\documentclass[traditabstract]{aa} % for a referee version ,referee
\usepackage{txfonts,graphicx}

\begin{document}

\title{The cosmic epoch dependence of environmental effects on size evolution of
red-sequence early-type galaxies
\thanks{Table 1 is only available in electronic form
at the CDS via anonymous ftp to cdsarc.u-strasbg.fr (130.79.128.5)
or via http://cdsweb.u-strasbg.fr/cgi-bin/qcat?J/A+A/}.}
\titlerunning{The epoch-dependent environmental effects of size evolution of 
early-type galaxies} 
\author{S. Andreon}
\authorrunning{Andreon S.}
\institute{
INAF--Osservatorio Astronomico di Brera, via Brera 28, 20121, Milano, Italy,
\email{stefano.andreon@brera.inaf.it} \\
\\ 
}
\date{Accepted ... Received ...}
\abstract{
This work aims to 
observationally investigate the history of size growth of early-type galaxies and how
the growth depends on cosmic epoch and the mass of the halo in which they are embedded.
We carried out a photometric and structural analysis in the rest-frame
$V$ band of a mass-selected ($\log M/M_\odot >10.7$) sample of 
red-sequence early-type galaxies with spectroscopic/grism redshift 
in the general field up to $z=2$ to complement a previous work
presenting an identical analysis but
in halos 100 times more massive and 1000 times denser. We homogeneously derived
sizes (effective radii) fully accounting for the multi-component nature
of galaxies and the common presence of isophote twists and ellipticity
gradients. By using these mass-selected samples, composed of 170
red-sequence early-type galaxies in the general field and
224 identically selected and analyzed in clusters, we isolate the  
effect on galaxy sizes
of the halo in which galaxies are embedded and its dependence on epoch.
We find that the $\log$ of the
galaxy size at a fixed stellar mass, $\log M/M_\odot= 11$, has increased
with epoch at a rate twice as fast in the field than in cluster in the last
10 Gyr ($0.26\pm0.03$ versus $0.13\pm0.02$ dex per unit redshift). 
Red-sequence early-type galaxies in the general field reached the size of
their cousins in denser environment by $z=0.25\pm0.13$ in spite of 
being three times smaller at $z\sim2$. 
Data point toward a model where size growth is epoch-independent
(i.e., $\partial \log r_e /\partial z = c$), but with a rate $c$ depending on environment,
$\partial c /\partial \log M_{halo} \approx 0.05$. 
Environment determines the
growth rate ($d \log r_e / dz$) at all redshifts, indicating an external 
origin
for the galaxy growth without any clear epoch where
it ceases to have an effect. 
The larger size of 
early-type galaxies in
massive halos at high redshift indicates that their size grew buildup earlier (at $z>2$)
at an accelerated rate, slowing down at some still
unidentified $z>2$ redshift.
Instead, the size growth rate of red-sequence early-type galaxies 
in low-mass halos is reversed: it proceeds at an increased rate at late epochs
after an early period ($z>2$) of reduced growth, in agreement with the qualitative
hierarchical picture of galaxy evolution.
We found similar values of scatter around the mass-size
relation independently of environment and epoch, indicating that the amount of 
dissipation in the system forming the observed galaxy does not vary greatly with 
epoch or environment.
}

\maketitle

\section{Introduction}

The origin and evolution of galaxies are among the most intriguing and complex chapters 
in the formation of cosmic structures (verbatim from Madau \& Dickinson 2014).
Understanding galaxy evolution requires controlling cosmic time, 
environment, mass, halo growth history, AGN activity, and much more because likely a
combination of these factors leads to the quenching of star formation and the emergence 
of the galaxy populations we see in the nearby Universe.
In turn, galaxies live in dynamical environments where halo mass
certainly plays a role in shaping their properties because
the physical processes altering the star formation
history and the size growth are likely fundamentally different for objects
close to the bottom of the halo potential well and for
those still orbiting in a much larger halo. In particular, does
halo mass affect the size evolution of massive early-type galaxies?
In a hierarchical galaxy formation model, halo mass assembly histories systematically
differ in different environments, with sub-halos aggregating earlier in denser environments
(e.g., Maulbetsch et al. 2006). Therefore, galaxy evolution
is accelerated in dense environments, while early-type galaxies
in less dense environments should catch up with their cousins in denser environments
having experienced an earlier size growth. Semi-analytic models
does not reproduce this expected behavior, however (see Sec~5.2). If instead secular processes
are responsible for the size growth, then the growth should be environmental-independent.

The most effective way to put forward the effect of the halo 
is to compare galaxies in their own halo versus galaxies in halos of other galaxies, i.e.,
centrals versus satellites.  Using an older terminology, this is a
comparison of field versus cluster galaxies at a given galaxy mass 
when we want a large mass contrast
between the accreting and primary halo.

Observational evidence about the environmental
effects on size are conflicting or inconclusive and largely focus on the mere
existence of a difference: some works suggest no environmental
dependency (e.g., Rettura et al. 2010; Maltby et al. 2010; Valentinuzzi et al.
2010; Kelkar et al. 2015; Huertas-Company et al. 2013; Allen et al. 2015; Saracco et al.
2017), some others 
claim larger sizes in dense environments (e.g., Delaye et al. 2014; Lani et al. 2013; 
Yoon et al. 2017)
or, in a few cases, suggest a reverse trend (e.g., Raichoor et al. 2011). In general,
environmental studies based on surveys lack sensitivity because they
do not include massive clusters (or, if present, they provide a minority of galaxies). 
Instead, environmental studies including clusters often lack 
sensitivity because of the limited sample size and/or redshift range probed, or
often rely on field samples heterogeneously selected and analyzed.  

Putting forward the effect of the halo is furthermore complicated by the
heterogeneous nature of galaxies in terms of: a) colors (e.g., Sandage \& Visvanathan 1978)
and therefore star formation histories (e.g., Larson et al. 1980); b) morphologies (e.g., Hubble 1926) 
and therefore structure evolution (Dressler 1980); and, likely, c) stellar mass assembly history
(e.g., Baugh et al. 1996). The heterogeneous nature of galaxies complicates the study of their
evolution because of the not completely disjoint classes and of the 
difficulty of replicating the same classification at all redshifts. 
For example, while in the local Universe many works adopt the Hubble sequence,
in high-redshift studies morphological classes may be replaced by 
large Sersic (1964) index,  
quiescence (e.g., Newman et al. 2012),
or massiveness.
Because of likely diverse evolutionary paths of the different classes
(e.g., Moresco et al. 2013), a non-homogeneous
selection at different redshifts is prone to systematics. 
Furthermore, even focusing on one single class may not suffice when
composed of galaxies having likely heterogenous histories such as
quiescent galaxies, known to be a composite population of truly passive galaxies,
dusty star-forming galaxies (Williams et al. 2009; Moresco et al. 2013)
and recently quenched galaxies (Carollo et al. 2013; Andreon et al. 2016).

To complicate the issue, galaxies are multi-component stellar systems 
(have arm, bars, bulges, disks, etc.), yet their half-light 
radii are almost always derived as if they were single systems (often 
fitting a single Sersic profile
to the azimuthally averaged radial profile) which is prone to systematics and
complicates the interpretation of the found trends. 

Finally, the considered redshift may matter: studying a fixed redshift only, or
a reduced range, may only reveal a part of the picture because halo mass may
be important at one cosmic time and negligible at another, leading to apparently
conflicting results.

In Andreon, Dong, \& Raichoor (2016, Paper I) we derived half-light radii for cluster galaxies
on the red sequence and of early-type morphology
in the rest-frame $V$ band of 224 galaxies
with $\log M/M_\odot \gtrsim 10.7$ at $0.02<z<1.80$.
The analysis was based on HST imaging for all $z>0.03$ galaxies (i.e., with sufficient
resolution) and allowed
galaxies to be multi-component. We want to repeat here a fully homogeneous selection
and analysis, but for field galaxies, to isolate the effect of the environment.
With the data derived in this paper, we not only identify whether the halo has
an effect on galaxy structure, but we also show how the halo influence
depends on epoch. By comparing halos with masses from a few to several $10^{14} M_\odot$
(clusters) to halos hosting galaxies with stellar mass of $10^{11} M_\odot$, and hence
total mass of $\approx 10^{12} M_\odot$ (van Huiter et al. 2011), we are comparing
halos differing by two orders of magnitude in mass and three orders of magnitude in (central)
density. A group versus field comparison would instead explore narrower ranges.
In such a comparison, galaxies of a fixed mass, say $\log M/M_{\odot}=11$, 
will be all satellite in clusters (no brightest cluster galaxy is so light in the studied
massive clusters), while almost all are central in the field (the few galaxies in
groups with brighter/more
massive galaxies have been removed in our study, see Sec.~2).

Throughout this paper, we assume $\Omega_M=0.3$, $\Omega_\Lambda=0.7$, 
and $H_0=70$ km s$^{-1}$ Mpc$^{-1}$. Magnitudes are in the AB system.
We use the 2003 version of Bruzual \& Charlot (2003) stellar population synthesis
models with solar metallicity and a Salpeter initial mass function (IMF).
We use stellar masses that count only the
mass in stars and their remnants.
For a single stellar population, or $\tau=0.1$ Gyr model, the 
evolution of the stellar mass between ages of 2 and 13 Gyr is about 5\%. Therefore,
comparisons (e.g., of radii) at a fixed present-day mass are degenerate with comparisons
with mass at the time of the observations (see Andreon et al. 2006 for a 
different situation).

\section{Data and sample selection}

In this paper we want to mirror what was done for cluster galaxies, namely
to derive sizes and masses for galaxies of early-type
morphology (ellipticals and lenticulars) at $z<2$ with $\log M/M_\odot \gtrsim 10.7$
and on the red sequence when measured on a filter pair bracketing the $4000$
\AA \ break.

\begin{figure}
\centerline{\includegraphics[width=7truecm]{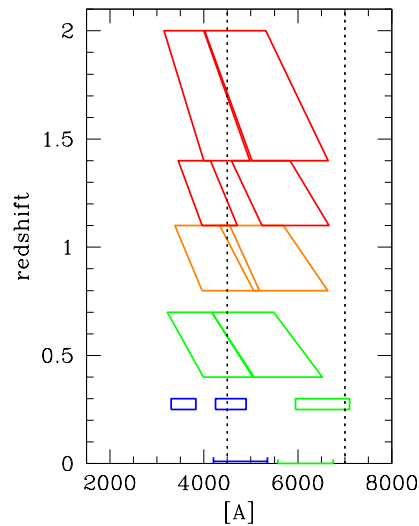}}
%\centerline{\psfig{figure=lambda_coverage_field.ps,width=7truecm,clip=}}
\caption[h]{Rest-frame wavelength coverage of the filters used for measuring
red sequence membership. The
two blue bands at $z\sim0.27$ are used to check the sensitivity of the
selection on the blue band used, see text.
}
\end{figure}

At $z>0.4$ we used GOODS-N and Hubble Legacy fields. At $0.25<z<0.3$ we used 
the larger (2 deg) COSMOS field, whereas at $25<D<42.9$ Mpc 
we used part of the SDSS DR12 (Alam et al. 2015). We excluded galaxies with $D<25$ Mpc
to minimize the effects of peculiar motions, while the other redshift ranges
were dictated by having an appropriate sampling in wavelength and resolution. 
In detail,
at $z>0.4$ for photometry and morphological classification we used 
deep Wide Field Camera 3 near-infrared (NIR) and 
Advanced Camera for Survey (ACS, Sirianni et al. 2005) 
wide field camera imaging of the 
Hubble Legacy Field (including the shallower and narrower GOODS-S),
distributed by Illingworth et al. (2016)
and  GOODS-N, distributed partly by CANDELS and partly by 3D-HST.
We run SExtractor (Bertin \& Arnouts 1996) in double
imaging mode, using F160W for detection at $z>0.8$ and F850LP at lower redshift.
Colors are based on fluxes within the detection isophote 
with a minor
correction for PSF differences across filters (derived 
in Paper I).
At $0.25<z<0.3$, we used 30-band matched photometry from
Laigle et al. (2015). Laigle et al. (2015)
gives photometry in the B, V, and R band rest-frame that we
used for measuring colors. These bands are interpolated from 
the available filters closest to them. 
For sizes and masses we used
instead HST F814W images.
At $25<D<42.9$ Mpc, we drew galaxies from the complete sample of early-type
galaxies in ATLAS3D (Cappellari et al. 2011, 2013).
We used the SDSS catalog for colors and
SDSS $r$ band images built from distributed frames for our own photometry, 
size measurements, and morphological classification.

Fig.~1 shows the 
rest-frame wavelength coverage of the filters used to determine
whether the galaxy belongs to the red sequence. 
At all $z$ we similarly sampled the $4000$
\AA \ break. 
At $0.25<z<0.3$ we used two blue bands 
to check the sensitivity to the adopted filter, and we found that the selected samples are
virtually identical. To measure the half-size radius we used
ACS or WFC3 images at $z>0.01$ ($F160W$, $F125W$, $F105W$, $F814W$, and $F814W$,
from high to low redshift).

We applied luminosity and color cuts:
we only considered galaxies brighter than a Bruzual \& Charlot (2013)
single stellar population
with $\log M/M_\odot = 10.7$ and $z_f=3$ because 
we are interested in $\log M/M_\odot \gtrsim 11$ galaxies. 
We then selected galaxies on the red sequence. 
We initially used a [-0.2,+0.2] mag range
around the red sequence (at $1.4<z<2$ and $0.25<z<0.3$), later reduced
to [-0.15,0.2] mag (other redshift ranges) because bluer galaxies turned out
to be late-type galaxies (but cost operator time).  
At $25<D<42.9$ Mpc, $\log M/M_\odot > 11.9$ galaxies
are overly represented in the parental (ATLAS3D) sample, while
they are absent in other samples. We therefore applied the additional
cut $\log M/M_\odot < 11.9$. 
Finally, we only retain elliptical
and lenticular galaxies and compute sizes in a band sampling about 
5000-6000 \AA \ rest-frame, as detailed in Sec.~3.

Since we are studying either NGC galaxies or medium-bright galaxies 
in famous fields, virtually all galaxies have
a spectroscopic redshift (or a distance measurement for galaxies in the very
nearby Universe). Spectroscopy comes
from grism/spectroscopic redshifts (and good photometry, \texttt{use\_phot=1}) 
listed in 3D-HST (Skelton et al. 2014) at $z>0.4$, Laigle et al. (2015)
at $0.25<z<0.3$, and from Cappellari et al. (2013 and references therein) at $25<D<42.9$ Mpc.

Concerning sample composition, we removed fainter galaxies of identified groups
because we only want central galaxies, plus the central one for a few
rich (crowded, to be precise) groups, both to widen the environmental range of our study  
and because the isophotal analysis in very crowded environments 
is unfeasible. As detailed in Appendix A, 
these bright and massive
galaxies carry almost no information on $\log M/M_\odot \sim 11$ galaxies,
which is the focus of this study, and therefore including or removing them
from the sample is irrelevant for the quantity of
our interest (independently of whether these galaxies
should be removed or kept in principle).

Our sample is virtually 
uncontaminated and almost complete, and incompleteness mostly random 
and therefore benign, as detailed in Appendix A.

Images are much deeper than needed and indeed morphological
classification and size measurements
of many of the same galaxies have already been performed in the past
using a reduced exposure time and down to fainter magnitudes 
(e.g., van der Wel et al. 2012; Cassata et al. 2011), but are recomputed here
using deeper observational material, with improved methods, 
and using a more uniform
sampling of the red band (used for size determination)
for homogeneity with the cluster sample.

\begin{center}
\begin{table}
\caption{Coordinates, masses, sizes, and PSF corrections.}
\begin{tabular}{l r r r r r}
\hline
\hline
ID & R.A. & Dec. & $\log M/M_\odot$ & $\log r_e$ & PSF corr \\
 & \multispan{2}{\hfill J2000 \hfill} & & [kpc] \\
\hline
\multispan{5}{$1.4<z<2.0$ \hfill} \\
12378 & 189.10034 & 62.15319 & 11.02 & -0.31 & -0.31 \\ 
14579 & 189.19055 & 62.16169 & 10.87 & -0.34 & -0.33 \\ 
17506 & 189.05850 & 62.17359 & 10.80 & 0.09 & -0.12 \\ 
...\\
\hline             
\multispan{5}{$25<D<42.9$ Mpc \hfill}\\ 
...\\
5854 & 226.94879 & 2.56856 & 10.87 & 0.22 & 0.00 \\ 
5864 & 227.38980 & 3.05274 & 10.95 & 0.30 & 0.00 \\ 
6278 & 255.20976 & 23.01096 & 11.14 & 0.30 & 0.00 \\ 
\hline            
\hline
\end{tabular}                                    
\hfill \break
Table 1 is entirely available in electronic form
at the CDS.
More digits than needed are reported for quantities.
\hfill \break     
\end{table}
\end{center}

\section{Morphology, size, and stellar mass}

As detailed in Paper I,
in order to derive effective radii and total luminosities (used later to derive
the galaxy mass) we fit the galaxy isophotes, 
precisely as done for galaxies in different environments at low and intermediate redshift
(e.g., Michard 1985; Poulain et al. 1992;
Michard \& Marshall 1993, 1994; Andreon 1994; Andreon et al. 1996, 1997a,b, etc.) and high
redshift (Paper I).
Briefly, isophotes are decomposed in ellipses plus Fourier coefficients
(Carter et al. 1978; Bender \& Moellenhoff 1987, Michard \& Simien 1988)
to describe deviations from the perfect elliptical shape. 
We classify galaxies by detecting morphological components in the
radial profiles of the isophote parameters.
Such a quantitative classification is more reproducible than morphologies
based on visual inspection (Andreon \& Davoust 1997) and
returns morphologies on average coincident
with those performed by morphologists such as Hubble, Sandage, 
de Vaucouleurs, and Dressler (Michard \& Marshall 1994; 
Andreon \& Davoust 1997).
By this morphological classification, we
remove from the sample non-early-type galaxies (i.e., spirals and irregulars),
only keeping elliptical and lenticular galaxies.

To compute the total galaxy flux, and from it the galaxy mass and size,
the flux between isophotes is integrated up to the last detected isophote,
in turn determining the curve of growth.
To extrapolate it to infinity, we fit the measured growth curve
with a library of growth curves measured for galaxies of different morphological
types in the nearby universe (de Vaucouleurs 1977), keeping the one that fits best.
The half-light isophote is, by definition, the isophote including half the 
total light. The half-light circularized radius, $r_e$, is defined 
as the square root of area included in the half-light
isophote divided by $\pi$. This definition allows us to 
define the half-light radius whatever the isophote shapes are and irrespective of 
whether galaxies have a single value of ellipticity and position angle, or
values that depend on radius,
as barred galaxies, lenticulars, and many ellipticals have. Our approach directly
addresses, and straightforwardly fix, the recently recognized problem represented
by objects not well represented by the idealized objects with
concentric ellipses of fixed
ellipticity and position angle and with perfect Sersic profiles, assumed instead in most
works, and for which a patch has been organized (Szomoru et al. 2010, 2013; Patel et al. 2017).

The background light is accounted for, and subtracted, by
fitting a low-order polynomial to the region surrounding the studied galaxy,
and accounting for the galaxy flux at large radii.
This also allows us to remove any
residual gradient present in the image, for example due to
scattered light. 

Masses of red-sequence early-type
galaxies are derived from $\lambda \approx 6000$ \AA \  luminosities 
assuming our standard BC03 SSP model with $z_f=3$ (which in turn
matches the red-sequence color) and checked for cluster galaxies in Paper I  
to introduce a negligible $0.10$ dex
scatter in mass and no bias
compared to a derivation based on fitting many photometric bands 
and $3000-6000$ \AA \ spectroscopy. Further checks are given in Sec.~4.1.3.

\begin{figure}
\centerline{\includegraphics[width=9truecm]{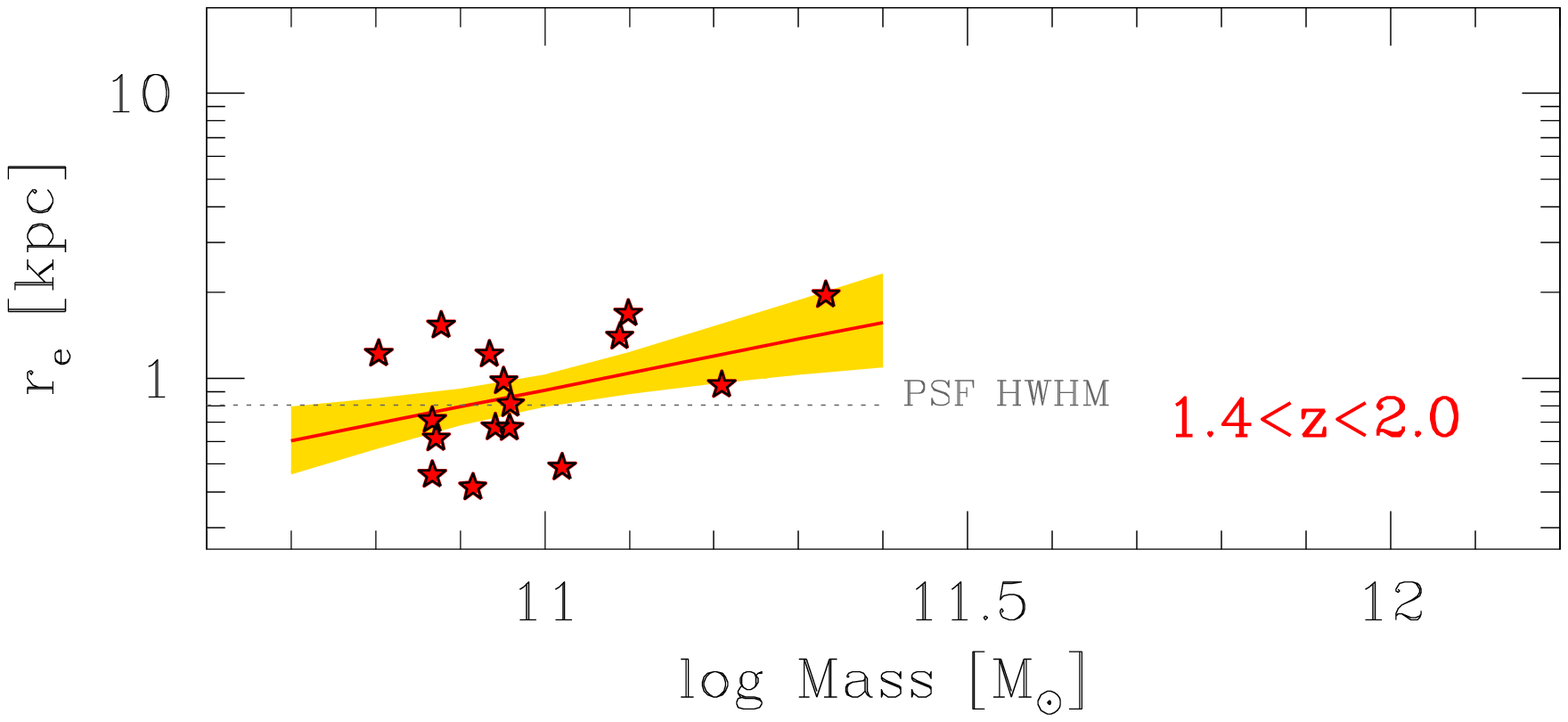}}
\centerline{\includegraphics[width=9truecm]{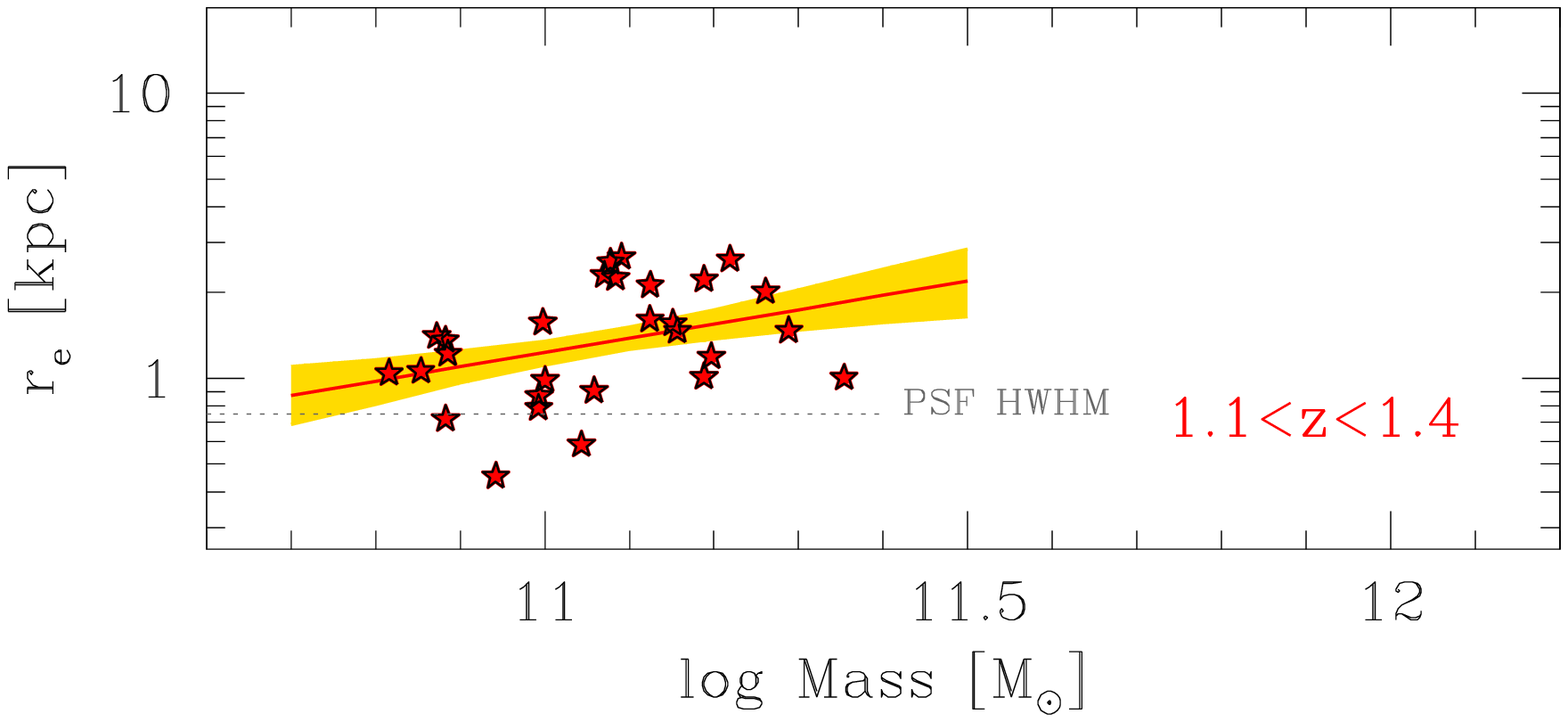}}
\centerline{\includegraphics[width=9truecm]{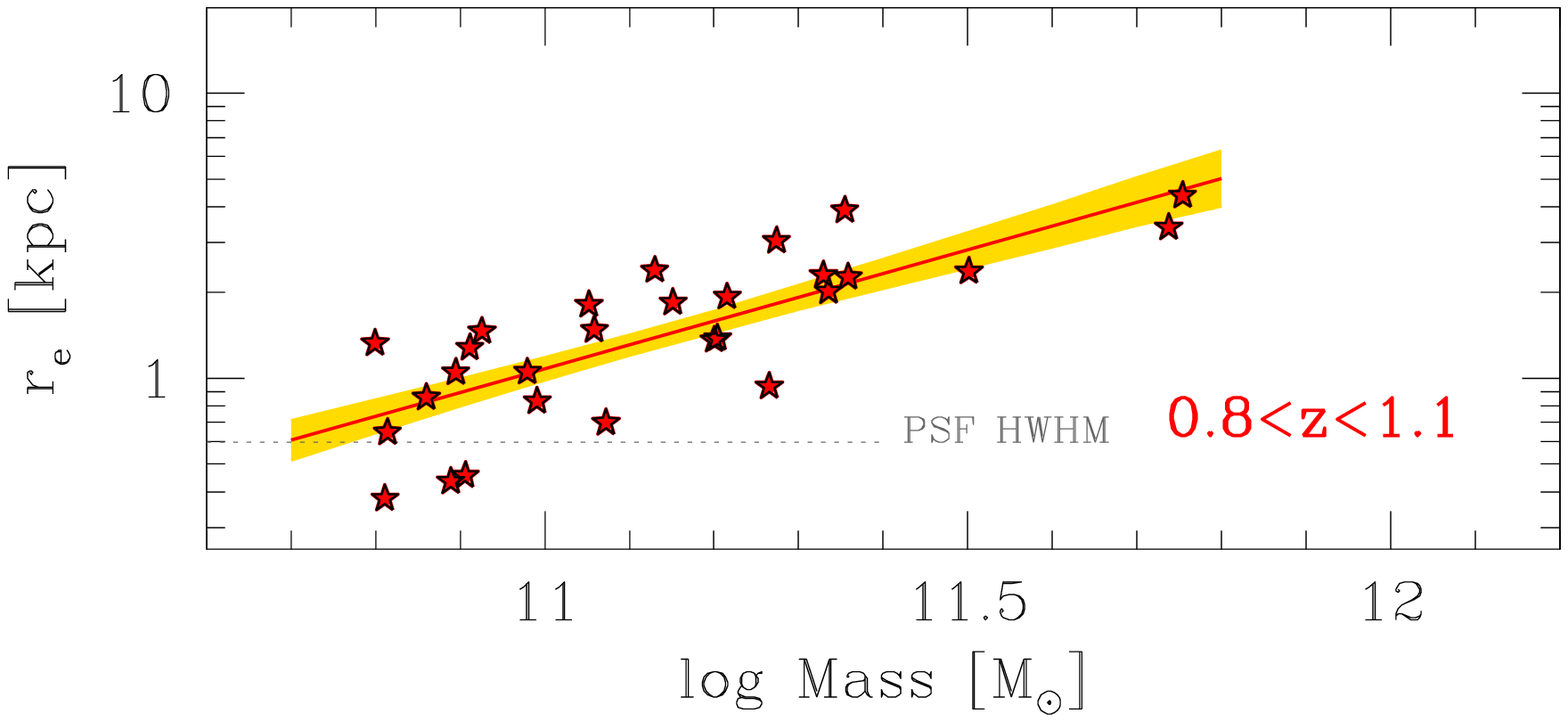}}
%\psfig{figure=mass_size_z16_field.ps,width=9truecm,clip=}}
%\centerline{\psfig{figure=mass_size_z122field.ps,width=9truecm,clip=}}
%\centerline{\psfig{figure=mass_size_z095field.ps,width=9truecm,clip=}}
\caption[h]{Mass-size relation of red-sequence early-type galaxies at $z>0.8$.
Sizes are corrected for PSF blurring effects.
The red solid line and yellow shading show the fitted mass-size
relation and its 68 \% uncertainty (posterior highest density interval).
The horizontal dotted line indicates the PSF half width at half maximum (HWHM).
}
\end{figure}

\begin{figure}
\centerline{\includegraphics[width=9truecm]{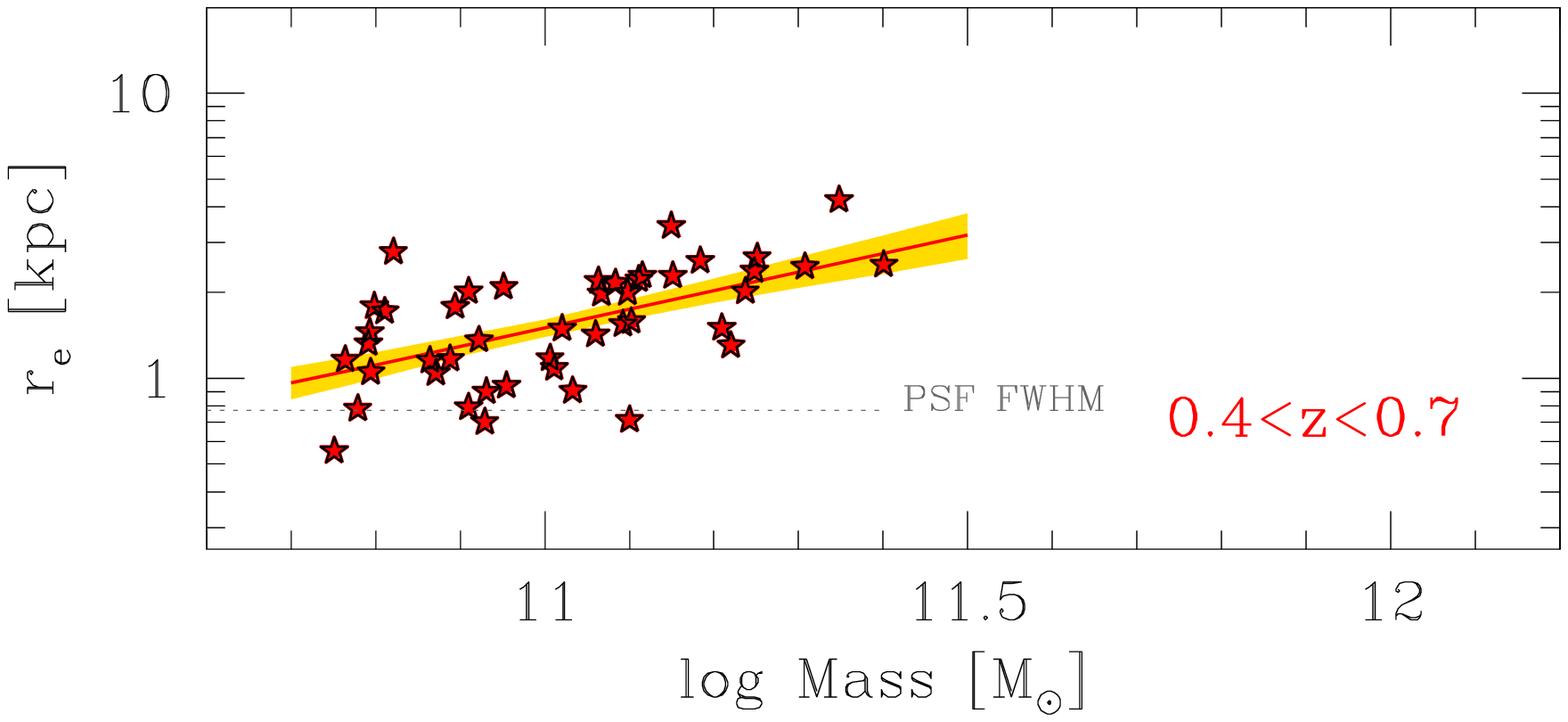}}
\centerline{\includegraphics[width=9truecm]{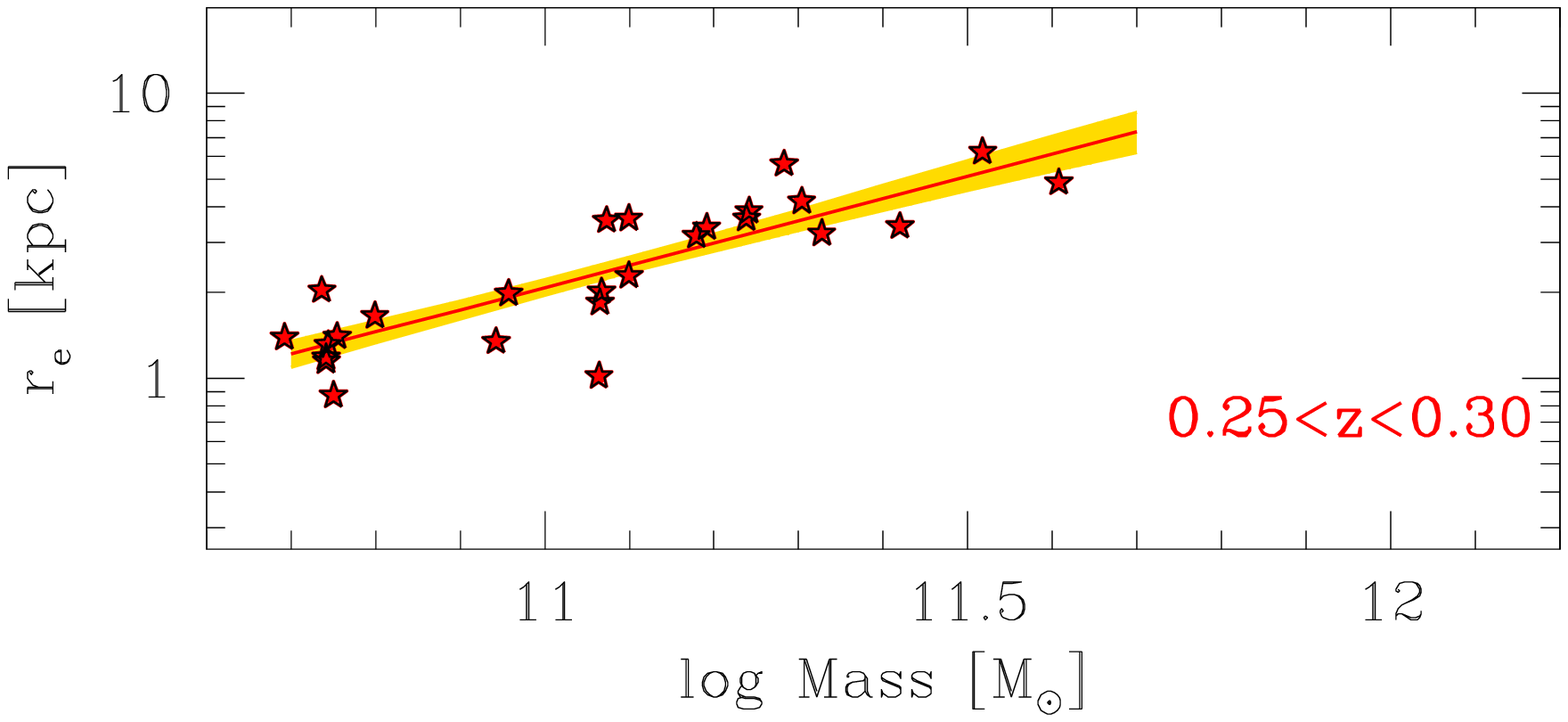}}
\centerline{\includegraphics[width=9truecm]{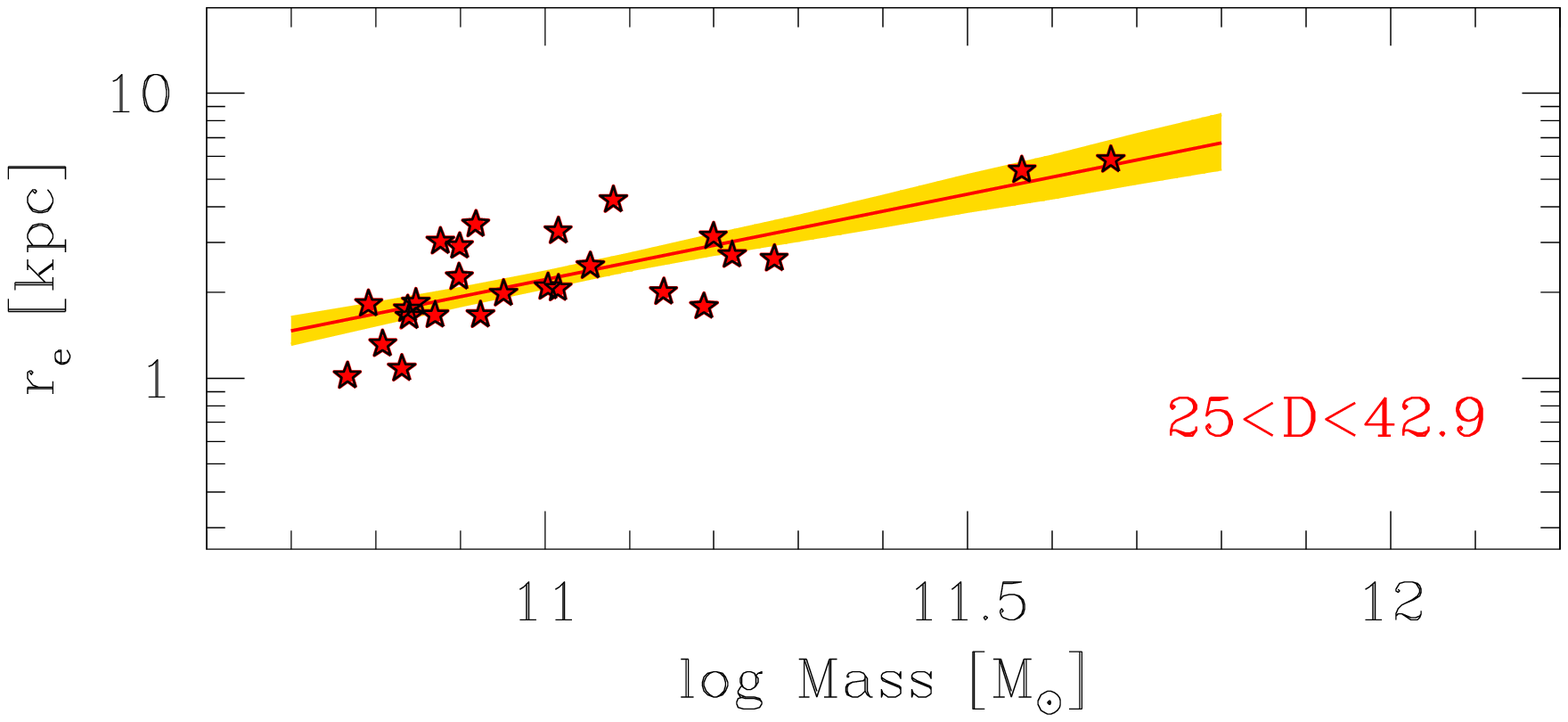}}
%\centerline{\psfig{figure=mass_size_z06field.ps,width=9truecm,clip=}}
%\centerline{\psfig{figure=mass_size_z027field.ps,width=9truecm,clip=}}
%\centerline{\psfig{figure=mass_size_z008field.ps,width=9truecm,clip=}}
\caption[h]{Mass-size relation of red-sequence early-type galaxies at $z<0.7$.
Sizes are corrected for PSF blurring effects, although the correction is
negligible.
The red solid line and yellow shading show the fitted mass-size
relation and its 68 \% uncertainty (posterior highest density interval).
The horizontal dotted line indicates the PSF HWHM (in the middle and bottom panels it 
is smaller than the displayed range).
}
\end{figure}

The PSF smears images and therefore makes galaxies appear larger
than they actually are.
We correct for PSF blurring by computing, following Saglia et al. (1993),
the size correction as a function of the observed half-light radius
expressed in FWHM units and assuming an $r^{1/4}$ radial profile. 
We applied the correction on a galaxy-by-galaxy basis, and we list
the applied correction in Table~1.
The correction is, in practice, zero at $z<1$, 
and then increases at higher redshifts mostly because of the broader PSF in NIR.
The correction is important only at $z>1.4$ because of the
reduced galaxy sizes there and the larger PSF. For cluster galaxies at the
same redshift and band (Paper I), the correction turned out to be negligible 
because of the larger galaxy sizes in richer environments.

\section{Results}

Table~1 lists coordinates, mass, and size (half-light radius) of the 170
early-type galaxies on the red sequence studied in this work. 
Fig.~2 and 3 show the mass-size relation of early-type galaxies on the red sequence
at the various redshifts. Identical plots are presented in Paper I for
cluster members.

\subsection{Checks}

\begin{figure*}
\centerline{\includegraphics[width=6truecm]{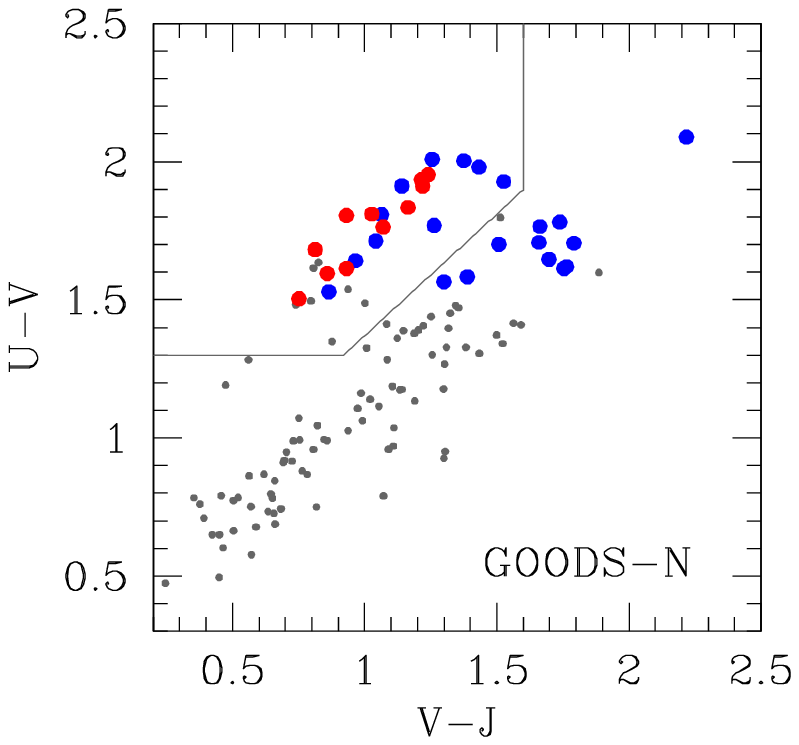}%
\includegraphics[width=6truecm]{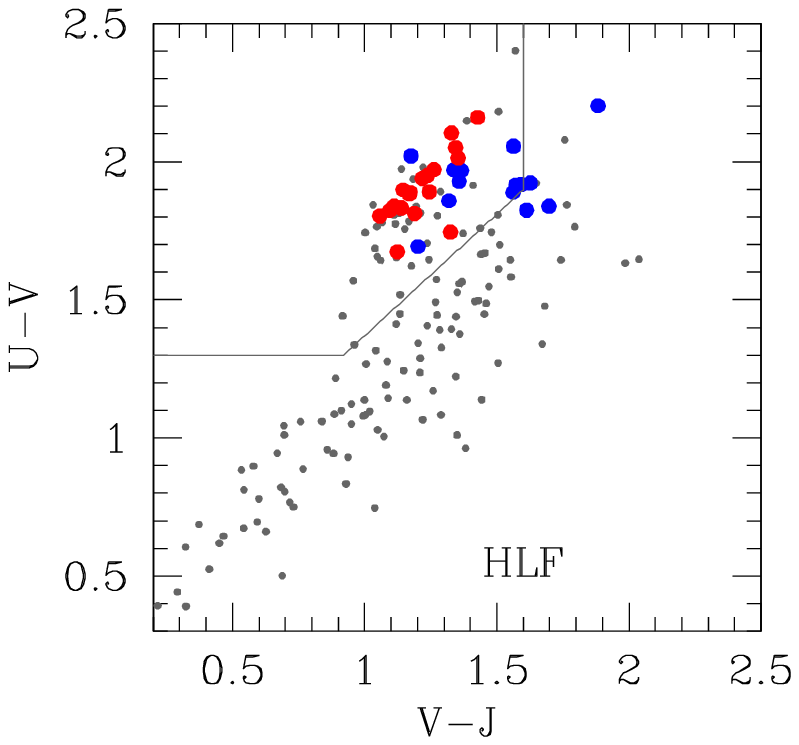}}
%\centerline{\psfig{figure=UVJ_GOODSN_z1420.ps,width=6truecm,clip=}%
%\psfig{figure=UVJ_HLF_z1114.ps,width=6truecm,clip=}}
\caption[h]{$U-V$ vs $V-J$ diagram of the galaxies with $H$ mag brighter than
a SSP with $\log M/M_\odot = 10.7$ and $z_f=3$, with good
photometry (use\_phot=1), $1.4<z<2.0$ in GOODS-N (left panel) or 
$1.1<z<1.4$ in HLF (right panel).
Red/blue
points are early-/late-type red-sequence galaxies with $H$-band derived masses 
higher than 10.7.
The gray solid line separates quiescent and star-forming galaxies according to
Williams et al. (2009). 
}
\end{figure*}

\subsubsection{Sample classification}

We classify galaxies following the
definitions of the morphological types. Other works do not apply this
morphological selection or adopt different
definitions for the morphological types leading in the case of clusters
to samples 30\% to 50\% contaminated by
non-early-type galaxies, as detailed in Appendix C of Paper I. 
More in general, UVJ quiescent galaxies are easily 30\% contaminated by
dusty star-forming galaxies (Williams et al. 2009; Moresco et al. 2013).
For the current field sample, we found that 
red-sequence
early-type galaxies are all UVJ quiescent galaxies (see Fig.~4; UVJ photometry and
classification is from Skelton et al. 2014), that 
red-sequence galaxies only compose 
two thirds of UVJ quiescent sample (see Fig.~5)
and that only about half of them are morphologically
early-type galaxies (see Fig.~5 and 6) in line with our previous works
on cluster galaxies (Andreon 1997; Paper I). Therefore, red-sequence
early-type galaxies compose just one-third of the UVJ quiescent population.
Figure~6 shows some illustrative examples of red-sequence UVJ quiescent galaxies yet
morphological late-type. The latter galaxies
have a morphological appearance showing that 
they are forming stars, or have just
stopped forming them, in spite of being called quiescent by UVJ colors and
being on the red sequence.
To summarize, UVJ quiescent galaxies are a broader population
than red-sequence morphologically early-type galaxies and the quiescent class
includes newcomers: one-third of them are yet not red enough
to be on the red sequence, and half of the
remaining are not yet morphologically early.

Galaxy populations selected with different criteria may well evolve differently
(e.g., Carollo et al. 2013;
Andreon et al. 2016).
Combining or comparing
samples selected in different ways is prone to systematics and must
be avoided.
We use consistent identical selection in color and morphology 
across environments and epochs.

\subsubsection{Half-size radius}

The half-light radius is the radius that encloses half of the
galaxy luminosity and our analysis strictly adopts this
definition. Many other works adopt a different definition of 
galaxy size, coincident with the half-light radius for ideal
galaxies rare in the real Universe (galaxies with a perfect Sersic
profile and without bulge, bar, disk, arms,
position angle twists, and without radial changes in the ellipticity).
As discussed in Appendix B of Paper I, these scale lengths 
should be combined with, or compared to, our half-light radii 
with great caution. 
At the light of the frequent presence
in real galaxies of isophote twists making a curved major axis, 
the advantages of major axis radii over circularized radii, 
proposed in some past works, should be
reconsidered when, as usual, major axis profiles are derived along
a single straight line that ignores the major axis curvature.

Restricting the attention to red-sequence morphologically early-type
galaxies only and for which half-light radius and scale lengths are measured in the same
photometric band (for which a rough agreement is expected), 
we found  $<0.1$ dex systematics with circularized 
scale radii in van der Wel et al. (2014), 
and 0.0 dex with those in van der Wel (2012).
For galaxies in our immediate neighborhood, our measurements agree with the 
values originally measured by Cappellari et al. (2013) and disagree with the values
listed in their table because the latter are scaled up by
$1.35$. We also have galaxies in common with Cassata et al. (2011), who use a 
redder band, however. They measured 0.07 dex
smaller effective radii %in a redder filter than we use, 
consistent
with expected color gradients of early-type galaxies.

\begin{figure}
\centerline{\includegraphics[width=5truecm]{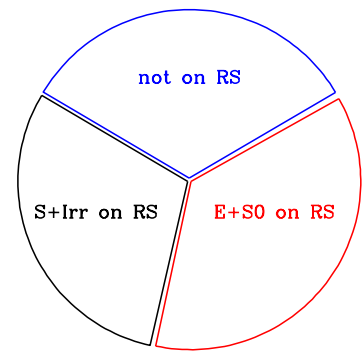}}
%\centerline{%\psfig{figure=piediag.ps,width=5truecm,clip=}}
\caption[h]{Partitioning of UVJ quiescent galaxies in galaxies not on the 
red-sequence (RS), RS galaxies of early-type
morphology (E+S0), and RS galaxies of late-type morphology 
(S+Irr). 
}
\end{figure}

\begin{figure}
\centerline{%
\includegraphics[width=3truecm]{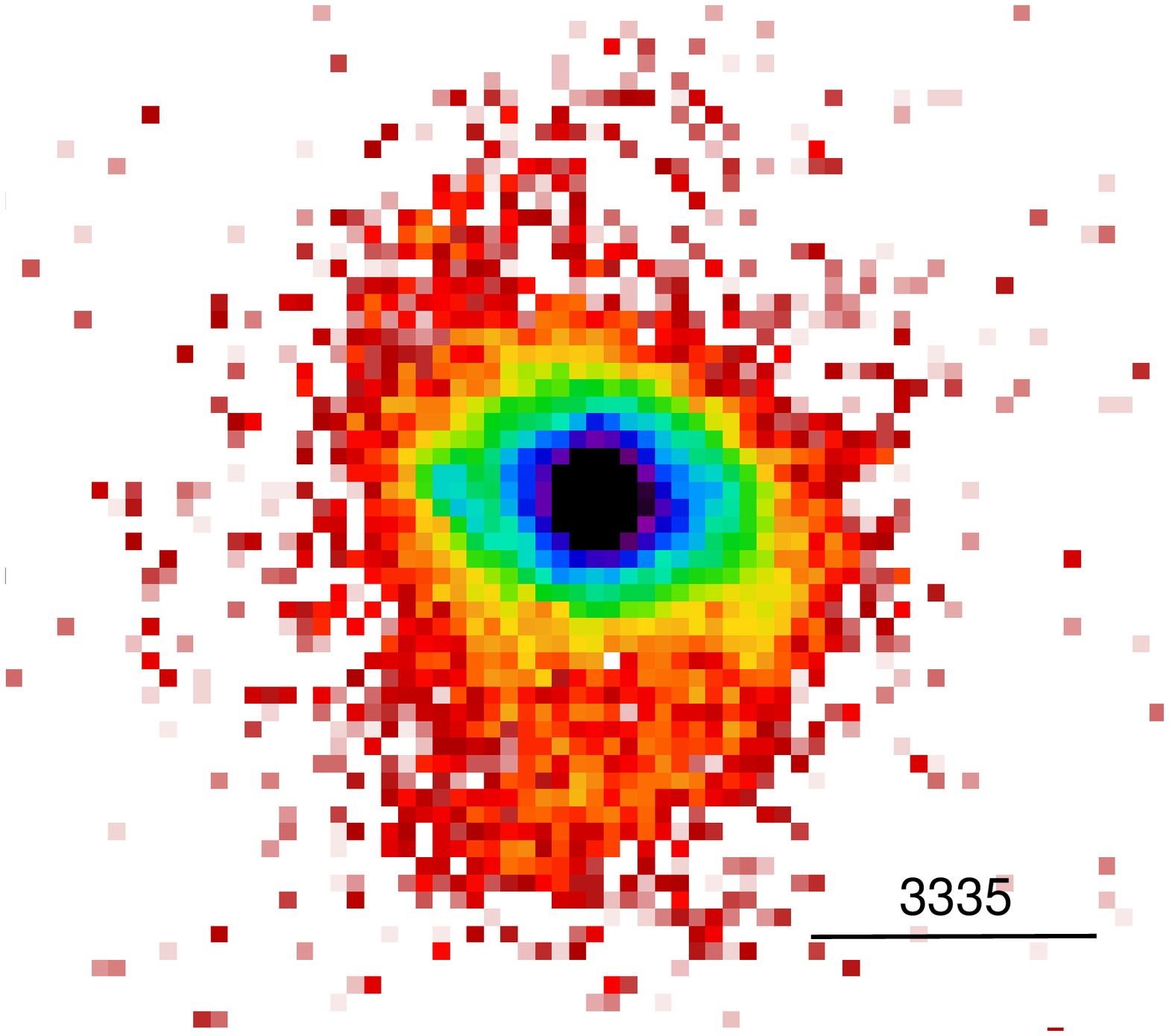}%
\includegraphics[width=3truecm]{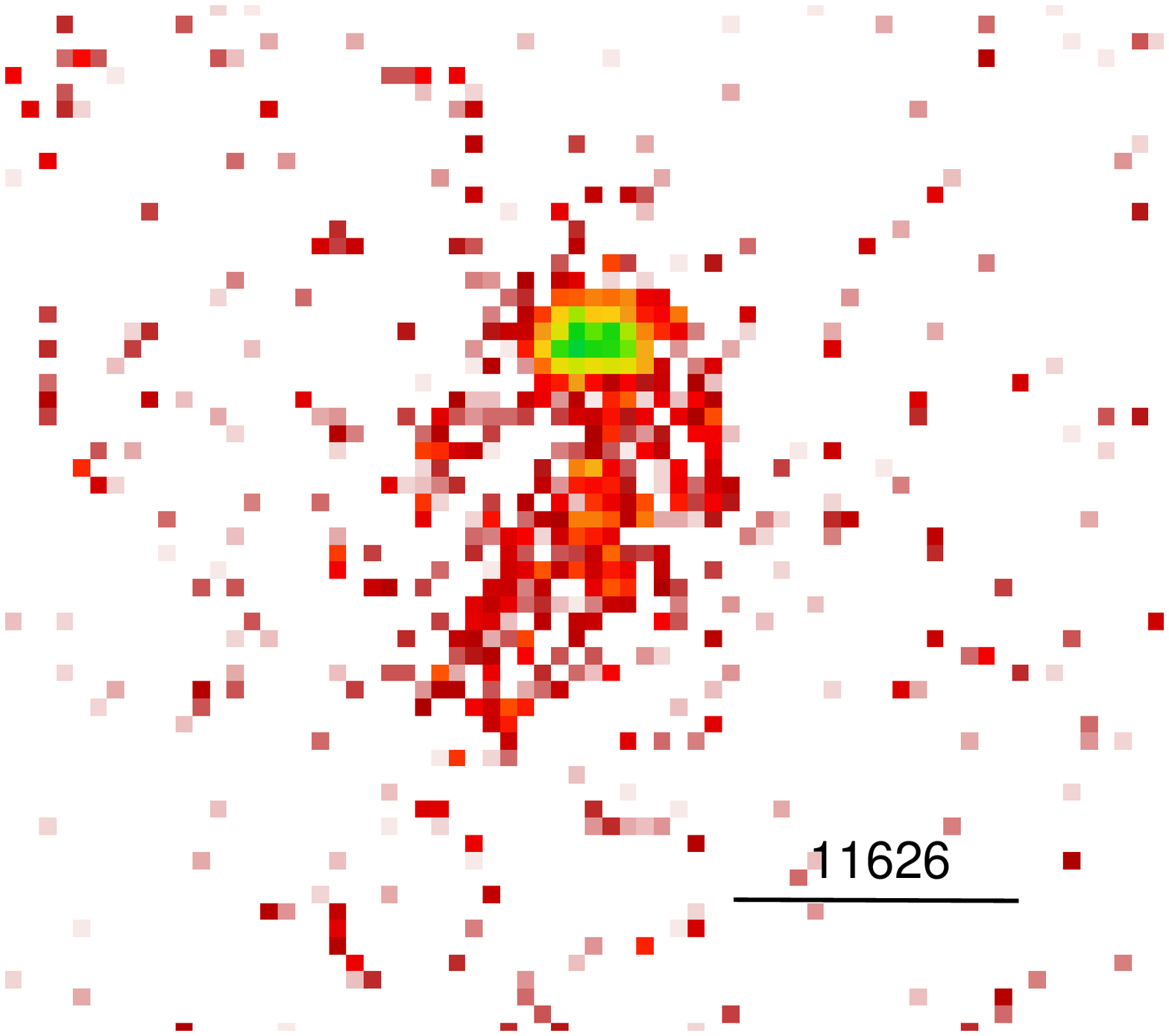}%
\includegraphics[width=3truecm]{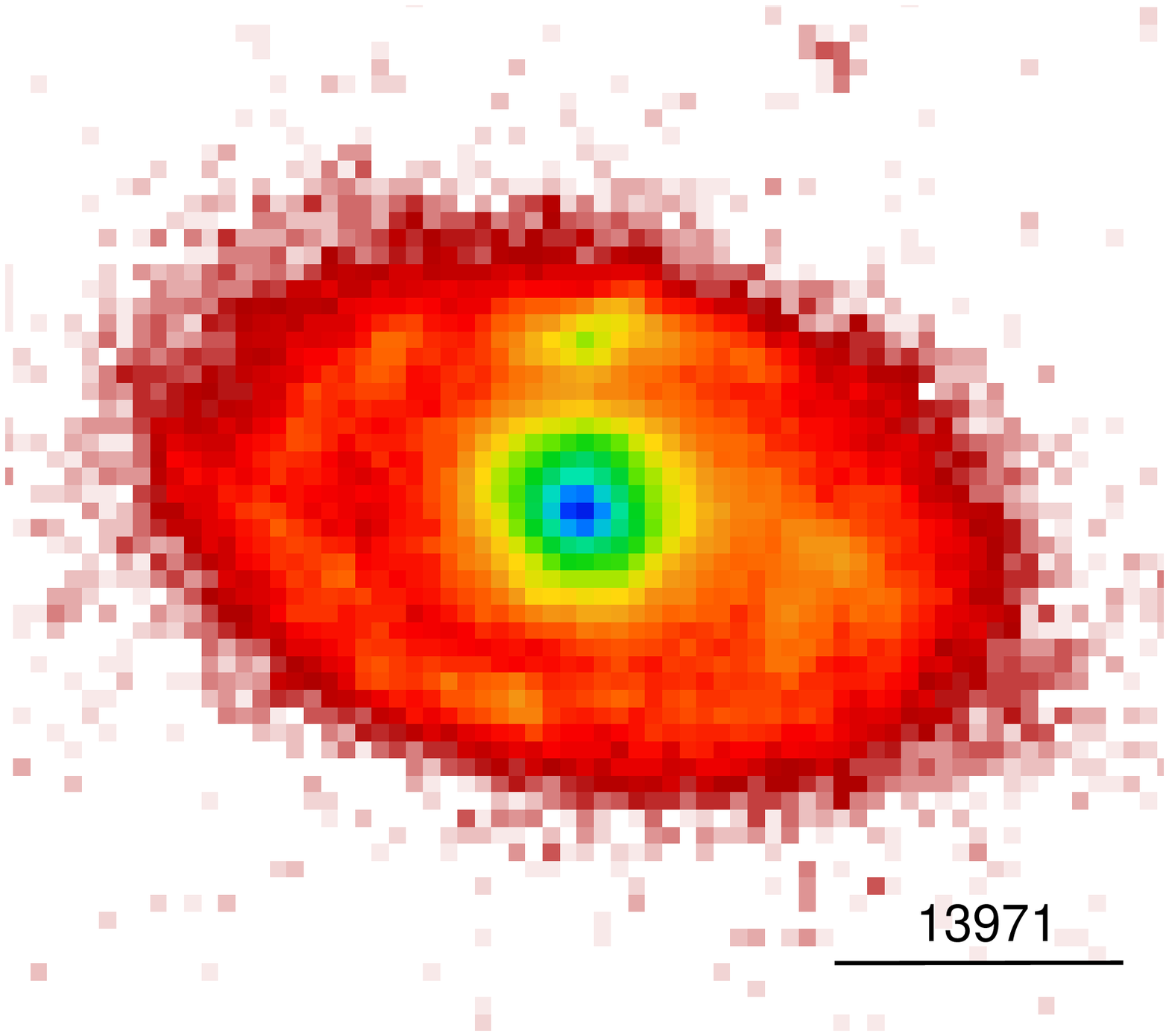}%
}
\centerline{%
\includegraphics[width=3truecm]{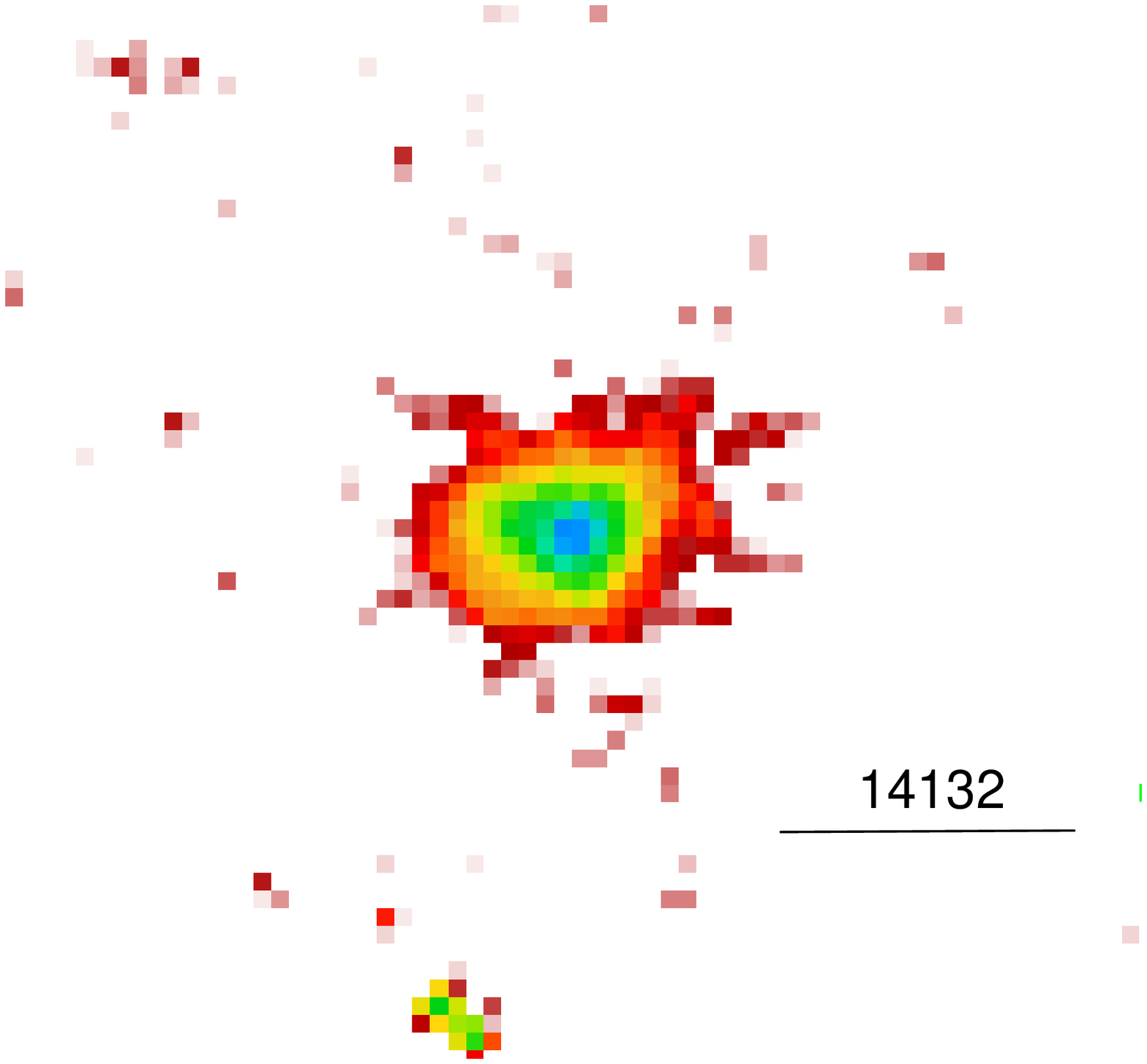}%
\includegraphics[width=3truecm]{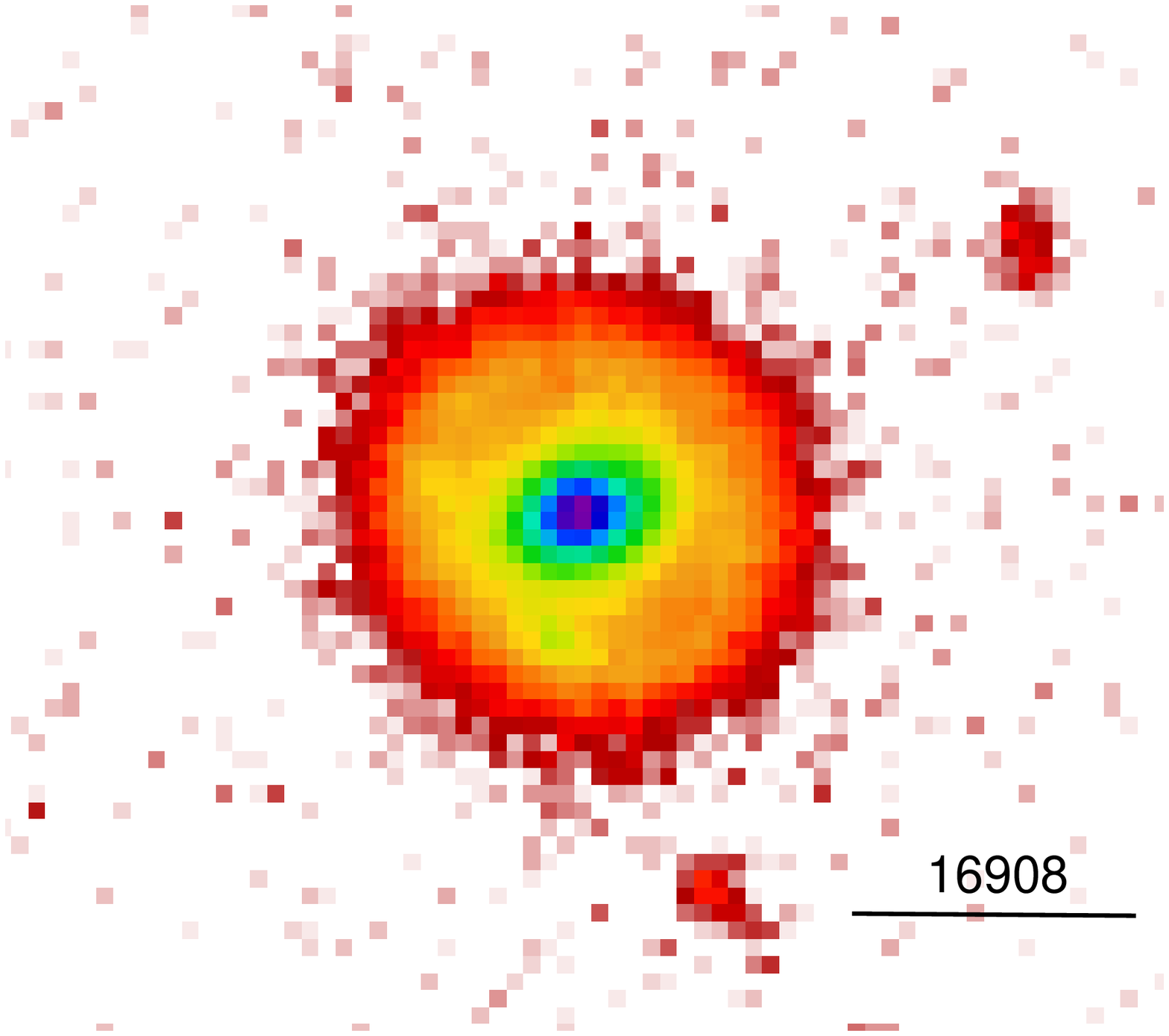}%
\includegraphics[width=3truecm]{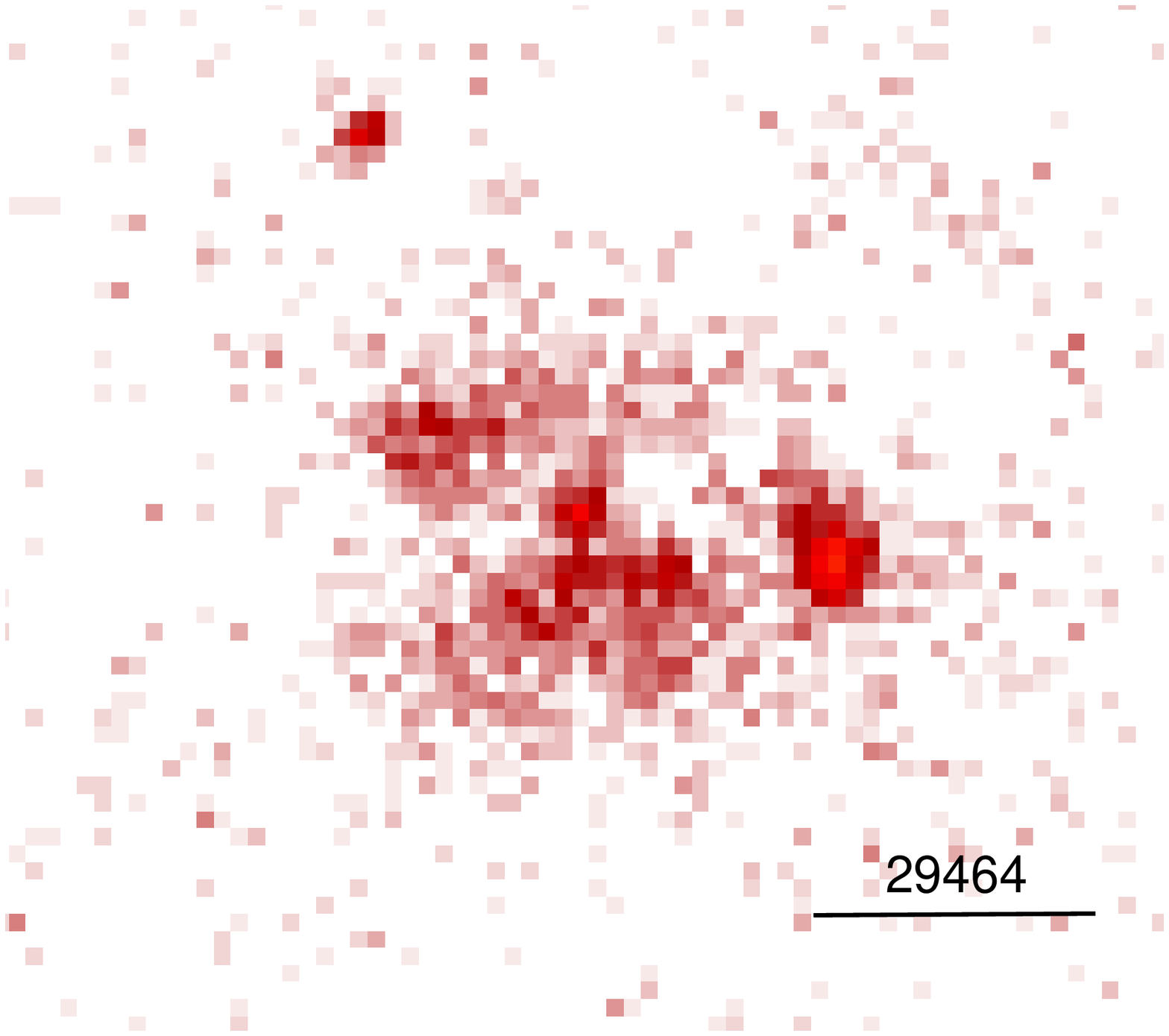}%
}
\caption[h]{Remarkable cases of UVJ quiescent galaxies on the red sequence
yet of late morphological type. These galaxies have manifestly
irregular or S-shaped isophotes. The tick is 1 arcsec. 
The numbers are the IDs in 3DHST paper.
}
\end{figure}

\subsubsection{Mass}

Our mass estimate, similar to those obtained from fitting photometric
data (e.g., Skelton et al. 2014, Laigle et al. 2016) 
comes from a total flux measurement (at $\lambda \sim 6000$ \AA \ in
our case) and a determination of the galaxy age (providing the $M/L$). Our total 
magnitudes agree
well with the total magnitudes of van der Wel et al. (2014) and
van der Wel (2012) for common
galaxies. The adopted spectral energy distribution template (a simple
stellar population with $z_f=3$, see Sec.~3)
matches the observed color of the red sequence and therefore it is not expected
to be grossly in error about the $M/L$.

However, after conversion to a common initial mass function, we agree
with the masses in Skelton et al. (2014) at high redshift, but we increasingly
disagree with decreasing redshift, up to 0.28 dex at $0.4<z<0.7$. We found
this to be due to different assumptions about the
galaxy ages: we adopted an old age ($z_f=3$), 
while  our red-sequence early-type
galaxies 
typically have a star formation time onset (usually called age) of 2 Gyr
independent of redshift according to the values tabulated
in Skelton et al. (2014) and adopted by these authors to estimate masses.
While at high redshift a 2 Gyr age roughly
corresponds to our assumed age,
at intermediate redshifts an age of 2 Gyr seems
implausible low (e.g., Thomas et al. 2005; Gallazzi et al. 2014). For example, 
massive galaxies 
(with no color or morphological pre-selection) at $z=0.7$ have
spectroscopic ages of 4.4 Gyrs (Gallazzi et al. 2014), while the typical age derived
by template  fitting of photometric data 
by Skelton et al. (2016) is 2 Gyr for the reddest objects (and less
for bluer ones). Since 
galaxies are younger in Skelton et al. (2016) than we assume (and increasingly so with
decreasing redshift), their mass is lower for the same
luminosity (and increasingly so with decreasing redshift), which explains the increasing
discrepancy with decreasing redshift. 
When our assumed age and that estimated in  Skelton et al. (2016)
are similar (at high redshift), masses turn out to agree. 

At $0.25<z<0.4$ our masses are larger by 0.17 dex than those Laigle et al. (2016)
estimate (from photometry) because the typical age of our red-sequence early-type
galaxies is 5.5 Gyr in Laigle et al. (2016) versus our adopted age of 8 Gyr.

Therefore, the adopted age has an important impact on the measured size  at
a given mass and on its evolution: a
0.3 dex discrepancy in mass measured at $z\sim0.6$ (and none at $z\gtrsim 1.2$)
with the Skelton et al. (2014) values,
and a size-mass slope of about 0.6 imply a systematic difference of 0.18 dex
in sizes (at low z only). This is larger than the error on the mean size of a 
$\log M/M_\odot=11$ and
comparable in absolute value 
to the variation we found in Sec.~4.2 between these redshifts. Therefore,
the correctness of the
derived size growths depends upon the accuracy of the assumed/derived
galaxy age.

\begin{figure}
\centerline{\includegraphics[width=9truecm]{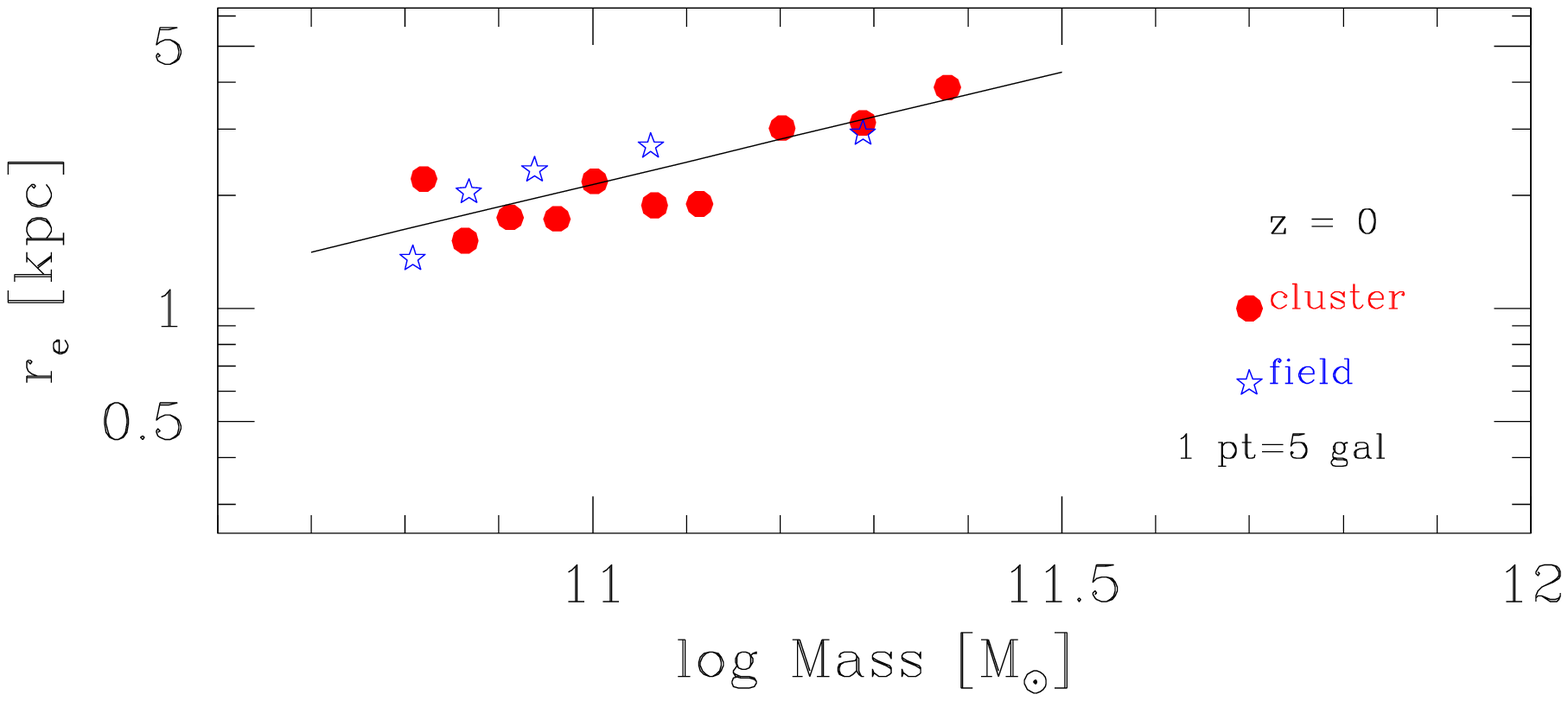}}
\centerline{\includegraphics[width=9truecm]{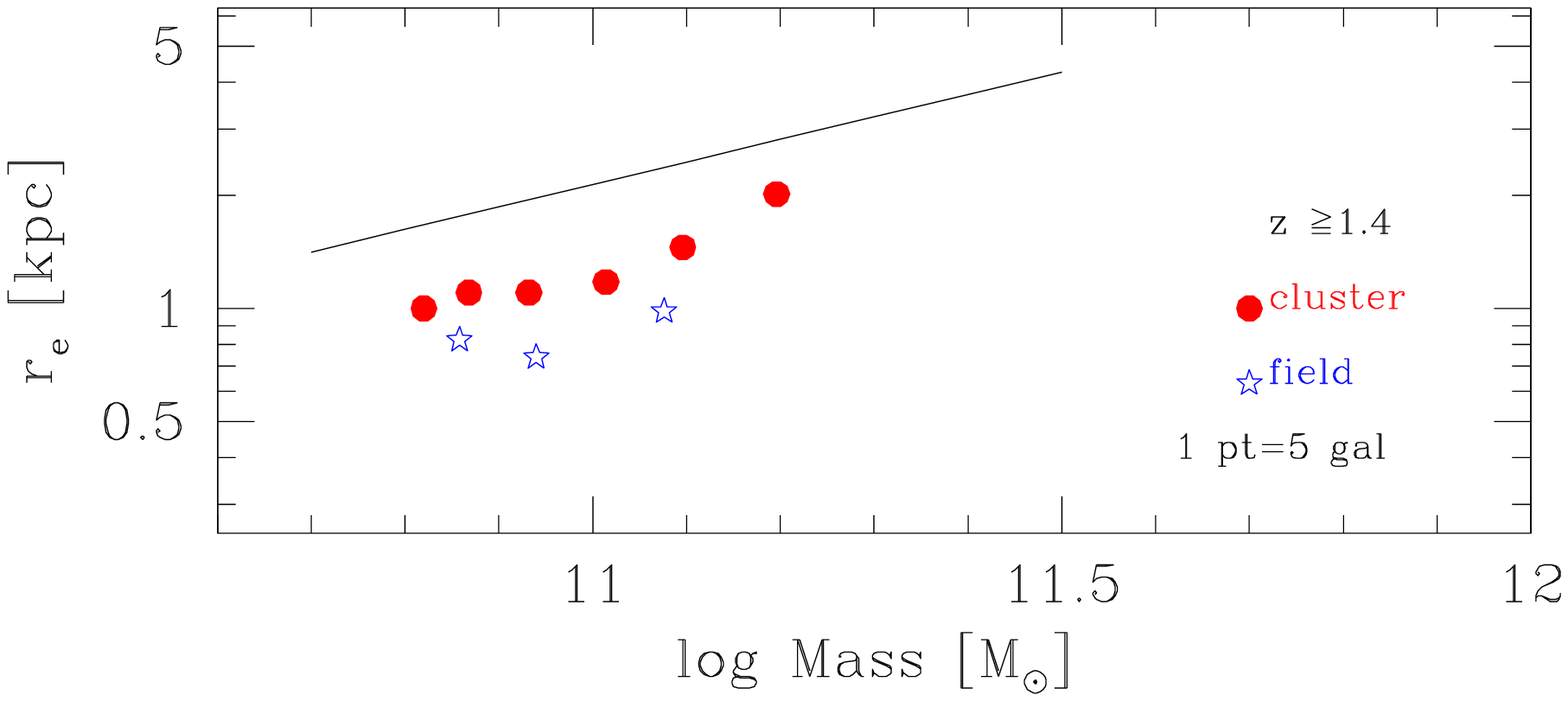}}
%\centerline{\psfig{figure=mass_size_stack_lowz.ps,width=9truecm,clip=}}
%\centerline{\psfig{figure=mass_size_stack_hiz.ps,width=9truecm,clip=}}%
\caption[h]{Qualitative comparison of the
mass-size distribution of $10.8<\log M/M_\odot<11.7$ red-sequence
early-type galaxies at
low (top panel) and high (bottom panel) redshift. Each point is the average of
five galaxies. The solid line indicates the scaling at $z=0$, and it is also shown in 
the bottom panel to show the evolution more clearly.
}
\end{figure}

\begin{figure*}
\centerline{\includegraphics[width=9truecm]{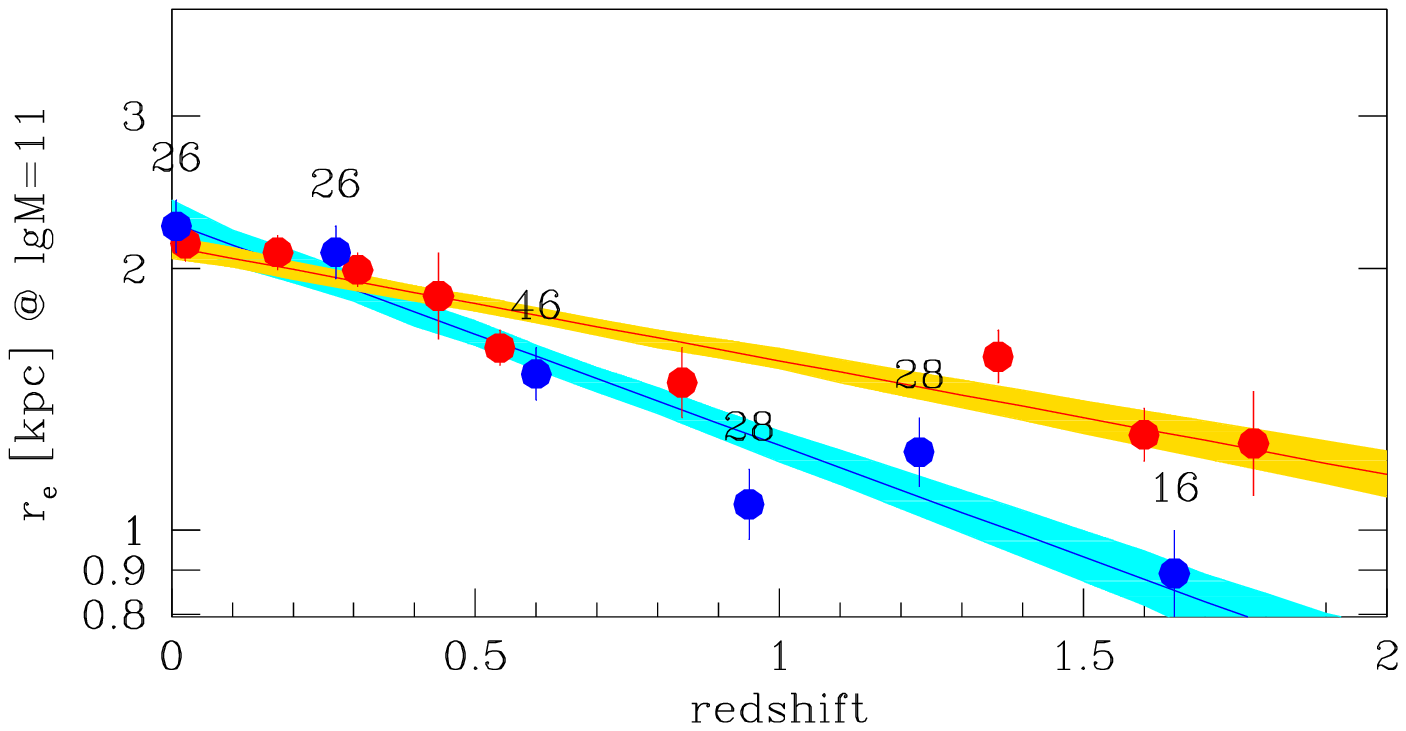}
\includegraphics[width=9truecm]{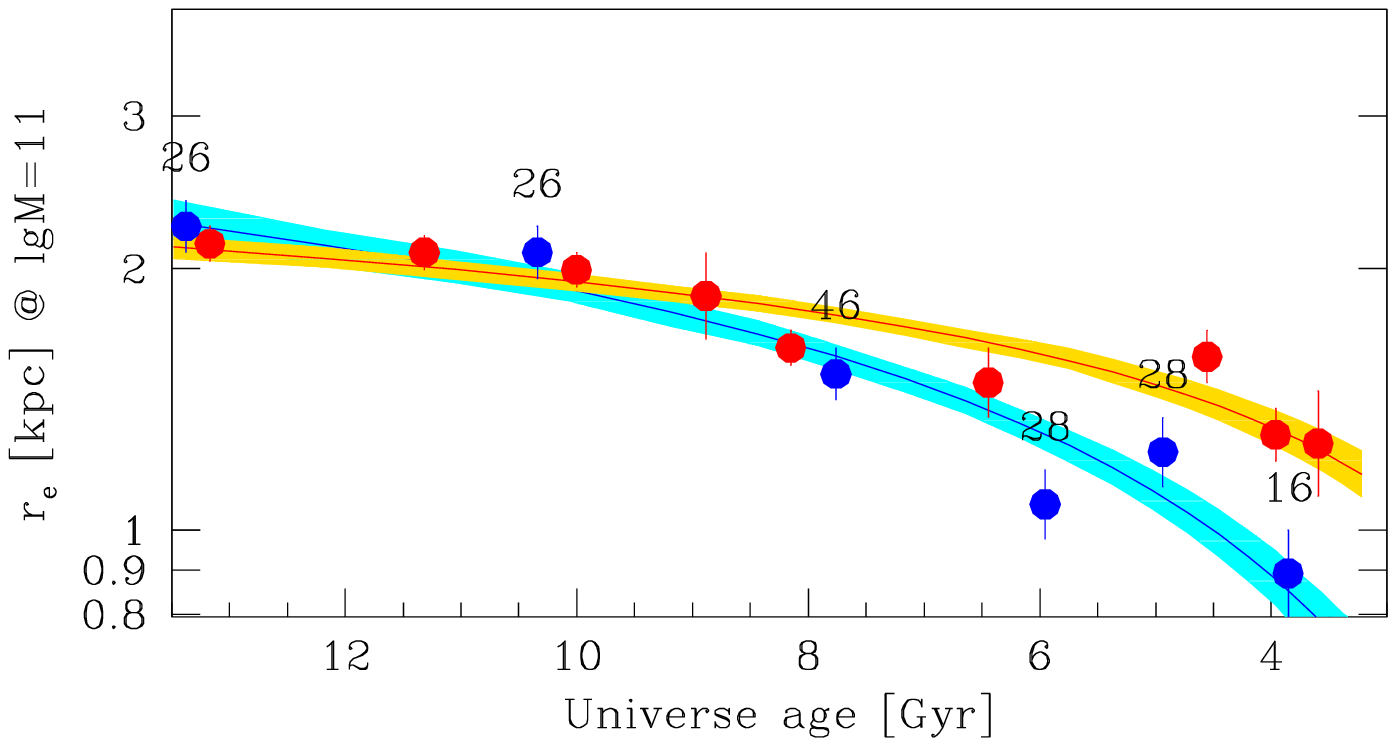}}
%\centerline{\psfig{figure=gamma_z.ps,width=9truecm,clip=}%
%\psfig{figure=gamma_t.ps,width=9truecm,clip=}}
\caption[h]{Halo effect on galaxy size and its dependence on look-back time.
The figure shows the mean galaxy size at $\log M/M_\odot=11$ vs redshift (left panel) or look-back time (right panel)
for red-sequence early-type galaxies in the field (blue points) and in cluster (red points). 
The number above the blue points indicates the number of combined galaxies. 
The solid line and shading show the fitted 
relation and its 68 \% uncertainty (posterior highest density interval).
}
\end{figure*}

\begin{table}
\caption{Mass-size fitting parameters: intercept $\gamma$, slope $\alpha$, and intrinsic
scatter $\sigma$ for the various samples}
\begin{tabular}{l r l l l}
\hline
Sample & \multicolumn{1}{c}{$\gamma$} & \multicolumn{1}{c}{$\alpha$} & \multicolumn{1}{c}{$\sigma$} & $N_{gal}$ \\  
\hline
$1.4<z<2.0$ &  $ -0.05\pm 0.05$ &  $  0.59\pm 0.37$ &  $ 0.21\pm 0.04$ &  $	  16$ \\
$1.1<z<1.4$ &  $ 0.09\pm 0.04$ &  $ 0.50\pm 0.26$ &  $ 0.20\pm 0.03$ &  $	  28$ \\ 
$0.8<z<1.1$ &  $ 0.03\pm 0.04$ &  $ 0.83\pm 0.14$ &  $ 0.19\pm 0.02$ &  $        28$ \\
$0.4<z<0.7$ &  $  0.18\pm 0.03$ &  $ 0.65\pm 0.15$ &  $ 0.17\pm 0.02$ &  $	  46$ \\
$0.25<z<0.30$ &  $ 0.32\pm 0.03$ &  $ 0.78\pm 0.11$ &  $ 0.14\pm 0.02$ &  $ 	  26$ \\ 
$25<D<42.9$ &  $ 0.35\pm 0.03$ &  $ 0.60\pm 0.12$ &  $ 0.14\pm 0.02$ &  $	  26$ \\  
\hline		   
\end{tabular}      				  
\end{table}

\subsection{Trends}

Qualitatively, the key result of this work is qualitatively illustrated in Fig.~7
using a portion of the data only,
where each point is the average of
five galaxies in order to emphasize the mean relation and downweight the scatter around it. 
The top panel shows that mass-size relations of the Coma cluster and of the local
field are very close to each other, while
galaxies at higher redshift (bottom panel) are smaller, and
those in sparse environments tend to be smaller than their counterparts in
clusters. In the following, we put on solid ground this
qualitative trend using the whole
dataset.

Half-light radii and masses in each redshift bin are fitted using a linear model with
intrinsic scatter $\sigma$ of the form
\begin{equation}
\log r_e = \gamma \ +\alpha (\log M/M_\odot -11) +\mathcal{N}(0, \sigma^2)
\end{equation}
adopting uniform priors  
for all parameters except the slope $\alpha$, for which we took instead a uniform prior 
on the angle $b=\arctan \alpha $. The parameter $\gamma$ is, by definition, the average size at 
$\log M/M_{\odot}=11$.  
Our approach improves upon
past works in two respects. 
First, the information
content of each individual point has a minimal floor given by the large scatter
around the mass-size relation (the scatter $\sigma$), 
while many past analyses (included all stacking ones) account only for the smaller
radius measurement error. 
Second, by leaving the slope free
we also allow different evolutions for galaxies of different mass,
discarded a priori by the works that keep the slope fixed (e.g., Carollo et al. 2013, 
or Yano et al. 2016). The free slope also de-weights galaxies with fairly different 
masses, allowing us to focus on $\log M/M_\odot =11$ galaxies.

Fig.~2 and 3 show the fitted trends and their uncertainty. Fit parameters are listed in 
Table~2. Slope and intrinsic scatter are consistent across redshifts, but
by leaving them free we do not overly constrain the fit negating a priori
mass-dependent evolutions and differences in scatter.

Fig.~8 shows the effective radius at $\log M/M_\odot =11$ (i.e., $\gamma$) as
a function of redshift. We fitted them with a linear relation in $z$, 
\begin{equation}
\gamma_z = \gamma_{11,z=0.6} \ +\beta (z -0.6)
\end{equation}
adopting uniform priors for the intercept at $z=0.6$, $\gamma_{11,z=0.6}$, and
a uniform prior on the angle $a=\arctan \beta $. In other terms, we are fitting
the effective radius at $\log M/M_\odot =11$ against $(z-0.6)^\beta$.

The mean size of red-sequence early-type galaxies in the field has grown
by $0.26$ dex per unit redshift in the last 10 Gyr (at fixed mass), see Table~3. Since the
growth is nearly linear in redshift and the relation between redshift and
look-back time is bent, this results in an accelerated evolution at earlier
epochs (Fig.~8,
right panel) in
agreement with previous works, for example with Newman et al. (2012, we found
an identical value of the slope $\beta$) based, however, on a broader class of galaxies 
(UVJ selected, see Sec.~4.1.1) and on scale lengths (see Sec.~4.1.2).

Fig.~8 also shows the effective radius at $\log M/M_\odot =11$ (i.e., $\gamma$)
for cluster galaxies (from Paper I), identically selected
and analyzed. Fitting the cluster data with eq.~2  
gives an evolutionary rate that is twice lower
than for
identically selected and analyzed field galaxies, see Table 3, indicating that at
$z<2$ the growth is twice as slow in clusters than it is in sparse environments. 
The larger
size at $z>1.5$ of cluster galaxies implies that growth was accelerated at a
redshift outside the studied redshift, i.e., at $z>2$.
Both fits are acceptable (at better than 90\% confidence level) 
in a $\chi^2$ sense, as can also be appreciated by detailed inspection of 
Fig.~8.

\begin{table}
\caption{Size evolution fitting parameters: intercept at $z=0.6$ 
and $\log M/M_\odot=11$, $\gamma_{11,z=0.6}$, and evolutionary
term $\beta$ }
\begin{tabular}{l l l l l}
\hline
Sample & \multicolumn{1}{c}{$\gamma_{11,z=0.6}$} & \multicolumn{1}{c}{$\beta$} \\  
\hline
field   &  $0.20 \pm 0.01$ & $-0.26 \pm 0.03$ \\
cluster &  $0.25 \pm 0.01$ & $-0.13 \pm 0.02$ \\
\hline		   
\end{tabular}      				  
\end{table}

As mentioned in the introduction, galaxies in massive halos are expected to experience
accelerated size growth compared to galaxies in sparser environments,  although
theory is unable to provide a robust quantitative prediction. For example, the Illustris
simulations does not fit the $z=0$ mass-size scaling (Nelson et al. 2015), 
and the successor IllustrisTNG simulation output galaxies whose size is half the earlier
simulation (Pillepich et al. 2018), and does not offer predictions for 
galaxies of different morphological classes or in different environments, nor does it
predict the epoch-dependent growth, hence effectively precluding comparisons.
Semi-analytic models do not reproduce this
expected behavior (see Sec.~5.2). The quality
of our data and the wide redshift sampling allow us to quantify the
qualitative expectation and establish the halo
effect, and also to determine the dependence of the amplitude on look-back time,
as determined above.

The epoch at which red-sequence early-type galaxies in sparse environments
catch up with their cousins in richer environments can be easily inferred (the intersection
of the two fits in Fig.~8), it is just matter of performing a joint fit of both
cluster and field data with a unique 
intercept for the two datasets
at the crossing redshift $z_{catchup}$. By taking a uniform distribution
as prior of $z_{catchup}$, zeroed for unphysical values of redshift, and a uniform 
prior on the angles, the joint fit of both cluster and field data gives 
$z_{catchup}=0.25\pm0.13$.  The delayed growth of
galaxies in sparse environments, combined with their fast growth at $z<2$, makes
galaxies of the same size around $z_{catchup}=0.25\pm0.13$.

Our data allow us to 
establish whether galaxies in different environments have similar or different sizes,
and the approximate time when their sizes match.
To further improve the localization of the catch-up redshift,
a dataset that more densely samples the low-redshift Universe is needed and, furthermore,
a redshift-unbinned analysis is preferable (and easy to implement, for example as 
in Andreon 2012).

Fig.~9 shows that
the scatter around the mass-size relation, $0.15-0.20$ dex (Table 2, see Paper I for 
cluster values) is fairly constant with environment and epoch, with some
possible indication of a larger value in the field. The 
scatter measures the variability from galaxy to galaxy of the amount of 
dissipation, integrated over cosmic time,
in the system that will form the observed galaxy. Its non-zero value
indicates that there is some variation from galaxy to galaxy. 
The little or no evolution
seen in both field and cluster environments and the little or no
difference between their amplitudes in the two environments 
indicates that the amount of dissipation of the system that formed the
observed galaxy does not vary greatly with epoch or environment.

\begin{figure}
\centerline{\includegraphics[width=8truecm]{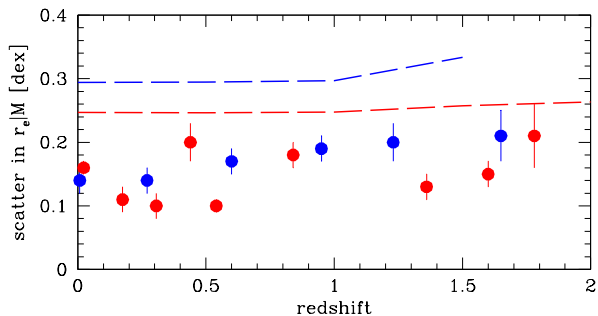}}
%\centerline{\psfig{figure=plotscat.ps,width=8truecm,clip=}}
\caption[h]{Scatter around the mass-size relation vs redshift. Red points
are cluster measurements, blue points field measurements, and red/blue
dashed curves are model predictions for cluster/field.  
}
\end{figure}

\section{Discussion}

\subsection{Comparison with other determinations}

When comparing our results with the results from other works, it should be 
remembered that red-sequence early-type galaxies
only are one-third of quiescent galaxies frequently studied in the literature,
and the broader class may evolve differently from each part, as already
pointed out.
Our red-sequence early-type galaxies form a more 
homogeneous and narrower
class than quiescent or Sersic-index selected samples. 
In addition, our derivation of the
half-light radius accounts for the common features of early-type galaxies, while
many works adopt scale lengths susceptible to the presence of galaxy 
morphological features. 

Generally speaking, when compared to other works
our analysis benefits from a larger redshift and environmental
baselines, allowing us to sample the epoch where environmental effects are more
manifest and at the same time, the epoch when such differences are less 
obvious. The larger redshift baseline allows us to study the epoch-dependence
of environmental effects, precluded to previous works. 

For example,
compared with the high-redshift work by Lani et al. (2013), our sample probes much 
wider redshift and environmental ranges, benefits from spectroscopic redshifts 
(i.e., is free of photo-z
catastrophic outliers), has the advantage of images that have
three times higher resolution 
(check in Fig.~2 the kpc scale for such degraded resolution), splits the galaxy
population into classes that are more homogeneous (their UVJ passive sample includes 
old early-type galaxies, and galaxies still star-forming or just quenched,
based on morphology, see Figs.~4 and 5), and size derivation allows 
galaxies to be multi-component. 
The larger redshift baseline allows us to study the epoch-dependence
of environmental effects, precluded
by their sample.

Compared with Cooper et al. (2012), the studied sample offers much wider 
redshift and environmental ranges, more bands 
for size determination to minimize
systematics due to color gradients (we used
three filters instead of one over the common redshift range), and
galaxy populations are split into more homogeneous classes.
The larger
redshift baseline allows us to study the epoch-dependence
of environmental effects, precluded by their sample.
As found by Delaye et al. (2014), we found larger galaxies in clusters; 
however our sample probes a much  
larger redshift range (their clusters are at $0.84<z<1.45$) allowing us to 
sample the catch-up redshift (not sampled by them). 
We also uses a homogeneous sampling of rest-frame wavelength for radii determination
(they noted that wavelength differences between 
cluster and field may affect their conclusions, as also remarked by Saracco et al. 2017). 
Furthermore, galaxy
populations are split into more homogeneous classes, and size derivation
allows galaxies to be multi-component.

Generally speaking, our results allow us to understand the variance in the
literature results if
they are applicable to scale lengths and to the larger class of quiescent galaxies.
For example,
Huertas-Company et al. (2013) find no environmental dependence, but they studied
$z<1$ only, a redshift range where differences are small (see Fig.~6), and even more
so given the restricted range of environments in their sample (they lack rich
clusters). Similarly, at $0.4<z<0.8$, Kelkar et al. (2015) and, at $0.2<z<0.7$, Morishita
et al. (2017) found
no environmental difference because they focused on an epoch when
environment and halo mass show a small difference at most.
At $0.1<z<0.15$, Yoon et al. (2017) found no environmental effects
for $\log M/M_\odot \approx 11$ galaxies, in agreement with our results, but their
studied redshift range is too small to yield the redshift-dependence we
detect and their study focuses on an epoch when environmental differences
are minor.

Instead, Saracco et al. (2017) found no environmental difference between the sizes of
cluster and field elliptical galaxies at $z\sim1.3$, while we found one
for ellipticals and lenticulars. However, apart from differences in
morphological composition (we include all lenticulars, while 
Saracco et al. (2017) only include those difficult to distinguish from
ellipticals, such as non-edge-on lenticulars),
their galaxy selection differs between
cluster and field because galaxies are color (red-sequence) selected for the cluster 
sample, while galaxies
of all colors are considered for the field. We instead performed the
same color selection in both cluster and field. Bluer quiescent galaxies tend to be
larger than redder ones (e.g., Carollo et al. 2013; Belli et al. 2015), and
indeed young early-type galaxies are larger than old ones (Saracco et al. 2009).
When blue early-type galaxies are 
included exclusively in the field sample, they increase the mean size in
this environment, hence reducing the difference between cluster and field. 
Finally, two of three of their field samples have
measurements (band used for
size determination) or selections (morphological classification and redshift range)
differing from the cluster sample.

\begin{figure}
\centerline{\includegraphics[width=8truecm]{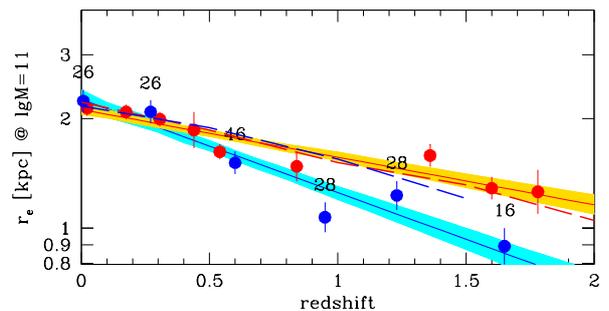}}
%\centerline{\psfig{figure=model_pred.ps,width=8truecm,clip=}}
\caption[h]{As the left panel of Figure 8, with superposed  
predictions (dashed curves) for galaxies in $\log M >14.3$ halos (red) and in $\log M <13.3$
halos (blue).
}
\end{figure}

\subsection{Comparison with semi-analytic models}

Fig.~10 compares the observed half-light size of $\log M/M_\odot =11$ early-type
galaxies on the red sequence at various redshifts (points and solid lines) with predictions
by the semi-analytic model in Shankar et al. (2013) for $10.8<\log M<11.2$
satellites
in massive halos (with $\log M >14.3$, red dashed line hardly distinguishable
from our fit to data) and for central galaxies of low-mass halos (with $\log M <13.3$,
blue dashed line). In simulations (and in observations too), satellites are both
galaxies having lost their sub-halo and those still having it but embedded
in a larger halo (type 1 and 2 in the code used by Shankar et al. 2013).
To better compare
with observations, simulations assume random statistical uncertainties 
in effective radius, stellar mass,
and halo mass of 0.08 dex, 0.1 dex, and 0.1 dex, respectively.
The Shankar et al. (2013) model
adopts the Guo et al. (2011) semi-analytic model inclusive of gas dissipation in major
gas-rich mergers and null orbital energies (parabolic orbits). 
In simulations galaxies are selected at the redshift $z$ 
to have a bulge-to-total
ratio higher than to 0.5 to mimic our morphological selection, and  a specific
star formation rate lower than $10^{-11}$ M$_\odot$/yr to mimic our 
observational red-sequence
selection (although the precise value of the threshold has virtually
no impact on model predictions). Since for observations we measure 
projected half-light radii, three-dimensional effective radii in semi-analytic
models are 
projected assuming a de Vaucouleurs profile, as in Shankar et al. (2013).
With the above choices, models predict a $(1+z)^{\sim 0.5}$ evolution
in both environments (Fig.~10).

The agreement between model predictions and observations 
is impressive for galaxies in massive halos at all redshifts (compare
the solid and dashed red curves) and even more so considering
that model sizes have not been re-normalized, unlike the comparison
in Huertas-Company et al. (2013). Once considering the lack of 
re-normalization, the agreement is also remarkable at $z\lesssim 0.5$
in sparse environments. In these sparse environments, the evolution
seen in the data over the whole redshift range
is closer to $(1+z)^{\sim 1}$ than to $(1+z)^{\sim 0.5}$ predicted
by the model. The original model by Guo et al. (2011) 
predicts similar trends in size evolution (F. Shankar, private communication).
Data support a {\it delayed} and faster size growth in low-mass halos compared 
to semi-analytic predictions: delayed because at $z\sim1$ simulations
overpredict galaxy sizes in the field and faster because the observed size growth 
proceeds at a faster rate (at $z<2$) in the data than in the model for field galaxies.
For cluster galaxies, the model growth rate is appropriate to $z\lesssim 2$
and any additional delay,
if any, should be minimal and fit in the short time available at $z>2$ in order not to
destroy the agreement between model and data
observed at $z<2$.

Fig.~9 compares the observed and predicted scatters around the
mass-size relation at various redshifts and for the two environments
(cluster/field are in red/blue, points are observed values, model predictions
are dashed lines). We note that the scatter in size induced by 
the scatter between true and estimated stellar mass and by 
size errors are consistently left inside the derived scatter
in size at a given mass, i.e., in the plotted points and curves.
The smallest and most precise observed scatter is about 0.1 dex, equal
to the expected combined effects of size and stellar mass errors (the
latter assumed for the model predictions).
The model predicts a close to constant scatter, as already
pointed out in Shankar et al. (2013) and 
as seen in the data, as welle as a larger scatter in the field environment,
as the data may also indicate. However, the model systematically 
overpredicts the observed scatter (as
already noted at redshift zero by Shankar et al. 2013), possibly 
indicating an overestimation 
of the amount of dissipation implemented in the model. We warn however
that different
methodologies are used to measure the scatter for the real data and the
semi-analytic data.

\subsection{Are environment effects on size growth epoch dependent?}

The continuity seen in the size growth in both environments (Fig.~8)
suggests that the
environment keeps its effects constant and continues to increase galaxy sizes at
different rates in different environments and that the environment never
stops having an effect on the galaxy sizes. The observed similarity of sizes
in different environments at low to intermediate redshift is due to this epoch corresponding
to the catch-up epoch, not to the cessation of the environment effects. The data seem
to suggest that there is
no transition epoch below which the environment stops affecting sizes (i.e., where
the derivative of trend with redshift becomes zero), but a catch-up epoch
at which the faster growth of galaxies in sparser environments makes them reach the size
of their cousins in more massive environments. As mentioned, because of the
non-linear relation between redshift and look-back time, a rate constant per unit
redshift is instead varying per unit time (Fig.~8).

We note that in the literature environmental effects are investigated comparing the effective
radius (at a given mass) in different environments at a single redshift
(or a small range). While useful, this choice is subject to degeneracies because the
observed size is the result of environmental effects integrated over time and there
may well be two different functions with 
identical integrals. For example, environment may play a major role, but at different times 
back in the galaxy history, hard to guess from a sample of equal-sized galaxies
in all environments at low or intermediate
redshifts: is their similarity the result of an environmental-independent growth,
or of two widely different growth histories having the same integral?  
The richness of our sample, and in particular the wide sampling in epoch allowing us
to measure the derivative of the galaxy size, breaks the degeneracy
of measurements at a fixed redshift
and allow to determine the environmental dependence, and its epoch dependence, 
of the galaxy growth.

\section{Conclusions}

We carried out a photometric and structural analysis in the rest-frame
$V$ band of a mass-selected ($\log M/M_\odot >10.7$) sample of 
red-sequence early-type galaxies with spectroscopic/grism redshift 
in the general field up to $z=2$. The sample is composed of 170
red-sequence early-type galaxies in the general field and complements
the sample of 224 early-type galaxies in clusters identically observed and
analyzed (presented in Paper I). The two samples are in
environments differing by 3 orders of magnitude in density and 2 orders of magnitude in
halo mass.

Because of the morphology and the narrower color selection of the sample
addressed in our study, 
red-sequence early-type galaxies are one-third only of the larger
quiescent galaxy population, the latter including bluer galaxies and
morphological late-type galaxies with evolutionary paths different from
the remaining part of the quiescent population. 
The tighter selections helps to disentangle the evolution
of the population from different levels of contaminations in different
environments or at different epochs, i.e., between galaxy properties
and sample selection.

We homogeneously, both across redshifts and environments, derived
sizes (effective radii) fully accounting for the multi-component nature
of galaxies and the common presence of isophote twists and ellipticity
gradients, allowing us to determine the epoch dependence of environmental
effects. Comparison with masses in the literature for common galaxies put forth
the important consequences on the inferred size evolution of systematics to
mass determination, such as the adopted age of the stellar population.

Observationally, red-sequence early-type galaxies in the field are smaller at 
high redshifts compared
to descendants and to objects at the same redshift in clusters, and their
size growth rate is about twice as large as than for objects in the cluster
environments ($0.26\pm0.03$ versus  $0.13\pm0.02$ dex per unit redshift) so
that objects in the field reached the dimension of those in
cluster at $z=0.15\pm0.12$.  
Environment affects early-type galaxy sizes in an epoch-independent way 
at $z<2$ when
the size growth rate is measured per unit redshift. In particular, 
there is no $z<2$ epoch when environment stops affecting galaxy sizes. 
Data point toward a model where size growth is epoch-independent
(i.e., $\partial \log r_e /\partial z = c$) but with a rate $c$ depending on environment,
$\partial c /\partial \log M_{halo} \approx 0.05 (=(0.26-0.13)/2.5)$,
where $2.5$ is the mass difference, on log scale, between field and cluster
halos (Sect.~1).

Early-type galaxies are larger in
massive halos at high redshift indicating that their size build up earlier (at $z>2$)
at an accelerated rate, slowing down at some still
unidentified $z>2$ redshift.
Instead, the size growth rate of red-sequence early-type galaxies 
in low-mass halos is reversed: it proceeds at an increased rate at late epochs
after an early period ($z>2$) of reduced growth, in agreement with the qualitative
hierarchical picture of galaxy evolution. Semi-analytical models considered
in this work get close to the observed behavior, but predicts a too fast early
growth and a too mild late evolution for galaxies in the field.

The scatter around the mass-size relation, $0.15-0.20$ dex, 
is fairly constant with environment and epoch and measure
the variety from galaxy to galaxy of the amount of dissipation
integrated over cosmic time
of the initial energy of the system that formed the observed galaxy. 
The little or no evolution
seen in both field and cluster samples and the little or no
difference between their amplitude in the two environments 
indicates that the amount of dissipation 
does not vary greatly with epoch or environment.

\begin{acknowledgements}
SA thanks Francesco Shankar for providing us with his predictions, 
CANDELS and 3D-HST teams for making HST combined images
available to the community,  and Charles Romero for useful discussions.
This work is based on observations made by the 
CANDELS Multi-Cycle Treasury Program and by the 3D-HST Treasury
Program with the NASA/ESA HST,
which is operated by the Association of Universities for 
Research in Astronomy, Inc., under NASA contract NAS5-26555.
\end{acknowledgements}

{}

\appendix

\section{Sample composition, completeness and contamination}

\begin{figure}

\centerline{\includegraphics[width=7truecm]{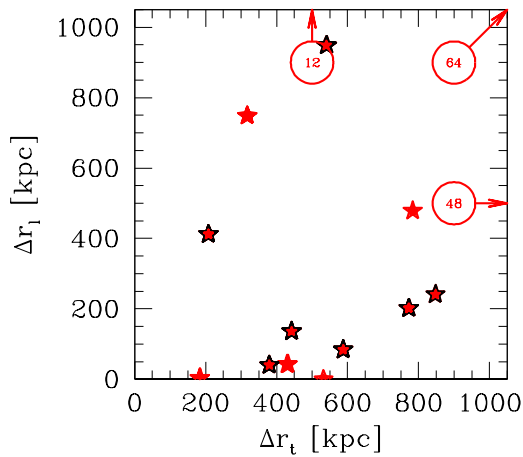}}
%\centerline{\psfig{figure=dr_dr.ps,width=7truecm,clip=}}
\caption[h]{Trasverse and longitudinal comoving distance from the nearest
galaxy in the sample. Symbols with black boundaries refer to GOODS-N, those
entirely red are galaxies in HLF. No galaxy in the COSMOS field falls in
the shown part of the plot. Circles with arrows indicate the number of galaxies
outside the plot. Galaxies in our immediate neighborhood are not considered in
this plot.
}
\end{figure}

In general,
random exclusion of galaxies from a sample does not bias the sample and results based on it 
(although they do reduce the sample power). Incompleteness instead biases results when 
galaxies are removed with preference (when their size is larger/smaller
than average for their mass, in the case of our work). 
Contamination may also bias results if unwanted galaxies
have properties different from those wanted (e.g., larger/smaller or more heterogeneous, at a
fixed mass, in the case of our work). Finally, differences in distribution in mass of the studied
sample do not bias results {\it at} a fixed mass, but can reduce the
statistical power of the sample. For example, adding $\log M/M_\odot > 11.9$ galaxies to a
sample of  $\log M/M_\odot \sim 11$ galaxies does not change the average size of 
$\log M/M_\odot \sim 11$ galaxies.

Our work is basically unbiased, but can suffer some limitations.
In our work some galaxies are randomly excluded from the sample for various
reasons, namely: a) very bright galaxies in the ATLAS3D are over-represented compared
to the other samples, especially those at high redshift. For this reason, we discarded all 
$\log M/M_\odot > 11.9$ galaxies and only consider a random sample among
those immediately less massive. These excluded galaxies 
have a much higher mass than those of interest and therefore
including or removing them does not changes the size of lighter galaxies that are
the focus of our work. Furthermore, we did not
select against galaxies larger (or smaller) than average for their mass, we
simply skip the analysis of bright ATLAS3D galaxies after having collected some of them in the sample;
b) target galaxies falling in unfortunate locations of images, including 
on the boundaries, on a sharp gradient
in exposure time, on a satellite track, on the top of a diffraction spike, or
close to bright star or an unrelated galaxy. As mentioned, since the reason for
exclusion is independent of the target galaxy (size at a given mass), exclusion is
random and our results
are not biased by the removal of these galaxies.

\begin{figure}
\centerline{\includegraphics[width=8truecm]{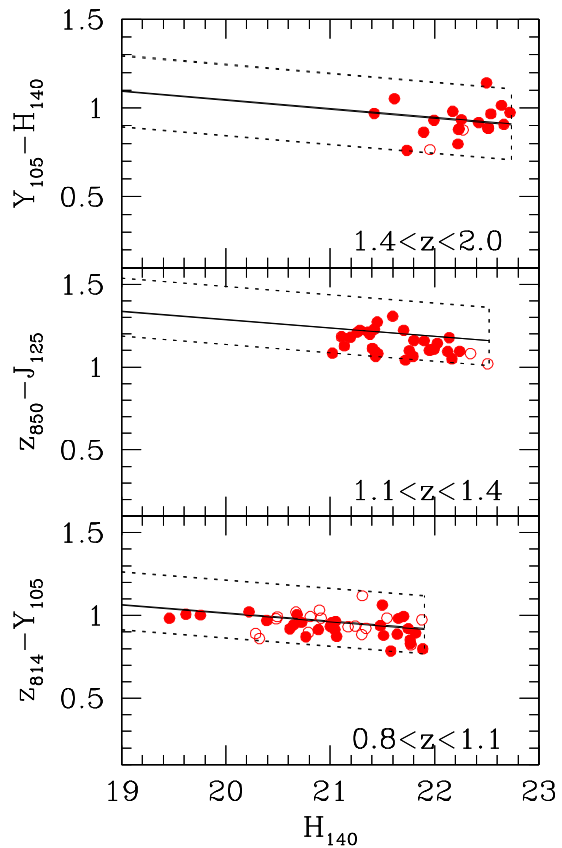}}
%\centerline{\psfig{figure=cmcol_hiz.ps,width=8truecm,clip=}}
\caption[h]{Color--magnitude plot of red-sequence early-type galaxies for galaxies at $z<0.8$. The slanted rectangles
indicate the selection region. Close points are analyzed galaxies, open points indicate
galaxies with unfeasible isophotal analysis. 
}
\end{figure}

About sample contamination, since we only want central galaxies, we removed 
from the sample all satellite galaxies in groups. This operation
was fairly straightforward because of the isolation of most of the galaxies and
the quality of the used data: Fig.~A.1 
shows the transversal and longitudinal comoving
distance of the galaxies left in the sample from the nearest galaxy in the sample
after removing obvious satellites plus a few
galaxies with unfeasible isophotal analysis (see below).
Groups with  $M_{500}\approx 10^{12} M_\odot$ have $r_{500}=100$ kpc (at $z\sim 1$, 
to be precise), and no galaxy pair is
that close in our sample. Even considering an unrealistic distance four times bigger
(in each direction, corresponding to an $M_{500}\sim 6\ 10^{13} M_\odot$ rich group 
or small cluster, hard
to miss in these images that are among the deepest ever taken), in our sample
there is one contaminating galaxy only. 
Therefore, even in the unrealistic scenario, our sample is 
at most contaminated by one galaxy, a $<1$\% contamination. 

\begin{figure}
\centerline{\includegraphics[width=8truecm]{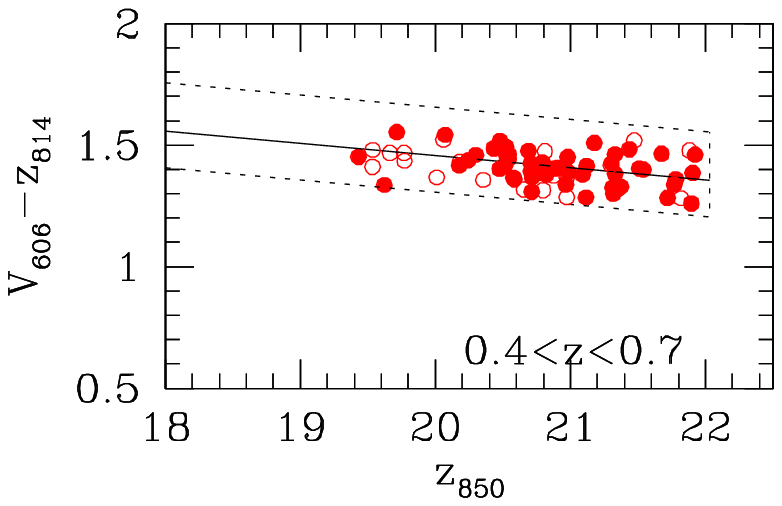}}
\centerline{\includegraphics[width=8truecm]{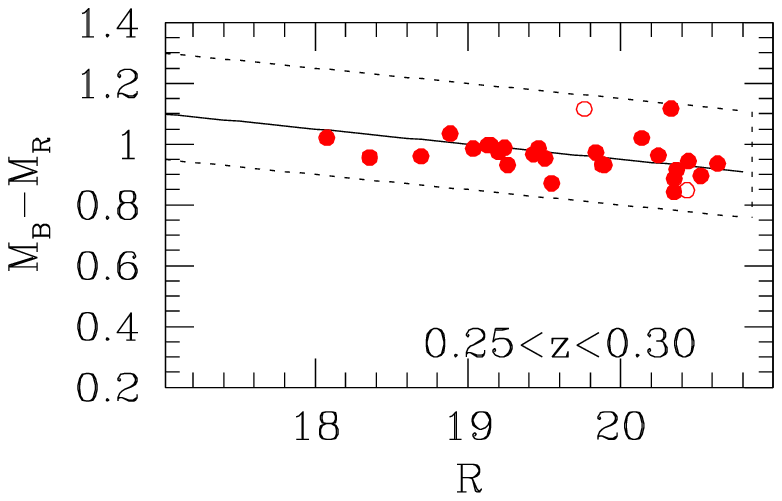}}
\centerline{\includegraphics[width=8truecm]{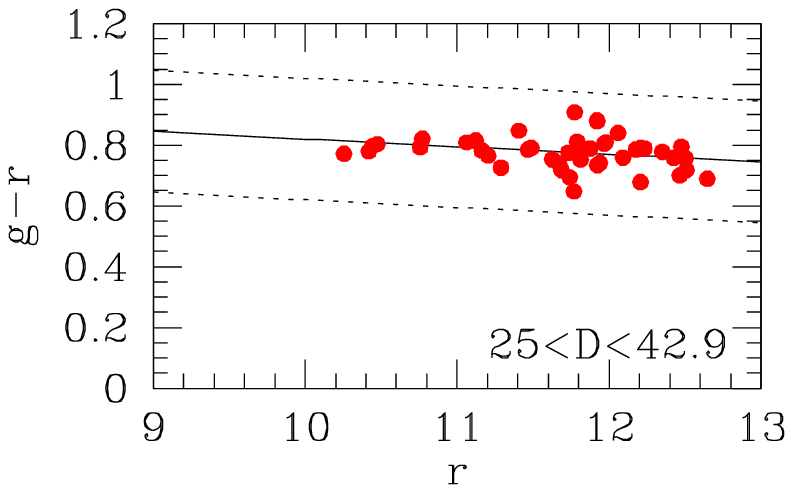}}
%\centerline{\psfig{figure=cmcol_z06.ps,width=8truecm,clip=}}
%\centerline{\psfig{figure=cmcol_cosmos.ps,width=8truecm,clip=}}
%\centerline{\psfig{figure=cmcol_atlas3D.ps,width=8truecm,clip=}}
\caption[h]{Color--magnitude plot of red-sequence early-type galaxies for galaxies at $z>0.8$. 
The slanted rectangles
indicate the selection region. Full Circles are analyzed galaxies, open circles indicate
galaxies with unfeasible isophotal analysis, and/or member of a group or, in the bottom
panel only, the sample randomly not selected for the isophotal analysis.
}
\end{figure}

We have two types of sample incompleteness. The first
is related to the removal of central galaxies in rich groups, operated
to widen the environmental range of our study and often as a matter of
necessity since these galaxies are often blended with their satellite and
therefore the isophotal analysis is unfeasible anyway. Since
the removal is performed by visual inspection, it is ambigous to some
extent. These galaxies
are often very bright and therefore carry almost no information on the size
of $\log M/M_\odot \sim 11$ galaxies; 
including or excluding them from the sample is irrelevant for the quantity of
our interest (and non dependently on whether these galaxies
should be in principle removed or kept). Second,
in a few cases isophote shapes or the galaxy environment turned out to be
too complex for our isophotal analysis to succeed. For example, 
our software is unable to deal with
isophote shapes that are not simply connected (i.e.,
with holes, such as for S0 with an important dust lane seen edge-on at 
high resolution). 
Our software is not able to deal with a
blend by, say, foreground spiral galaxies or multiple
faint galaxies on closeby lines of sight (if feasible at all).
The subsample of these galaxies
having $\log M/M_\odot \sim 11$ is the main source of incompleteness of our sample.
To bias our results, this subsample of missing galaxies should consist preferentially
of galaxies larger (or smaller) for their mass. 
Fig.~A.1 and A.2 show all studied galaxies (full circles) and the missed
galaxies (open circles). 
Only 15 \% of $\log M/M_\odot \sim 11$ galaxies
do not have a radius determination, and since the reason for
exclusion is almost independent on the target galaxy (we would have likely missed 
most of the galaxies if they were
slightly larger or smaller), exclusion is random (``ignorable" is the
appropriate statistical term, see the Gelman et al. 2004 book) and our results
are not biased by the removal of these few galaxies.

\end{document}